\documentclass[longauth]{aa} 

\newcommand{\Espresso}{\textsc{Espresso}}

\usepackage{subcaption}

\usepackage{graphicx}
\usepackage{xcolor}

\usepackage{txfonts}
\newcommand{\logg}{\ensuremath{\log g}}

\newcommand{\teff}{T$_{\rm eff}$}
\def\kms{\,$\mathrm{km\, s^{-1}}$}

\begin{document} 

\title{\Espresso{} observations  of  HE\,0107$-$5240 and  other CEMP-{no} stars with [Fe/H] $\le -4.5$  
\thanks{Based  on ESPRESSO Guaranteed Time Observations collected at the European Southern Observatory under ESO programmes 1102.C-0744, 0103.D-0700, 0104.D-0688, 1104.C-0350, 108.2268.001, P.I. P. Molaro.}
}
   
\author{D.~S. Aguado\inst{1,2} \thanks{E-mail: david.aguado@unifi.it}, 
P. Molaro\inst{3,4}, 
E. Caffau\inst{5}, 
J.~I. Gonz\'alez Hern\'andez\inst{6,7}, 
M. R. Zapatero Osorio\inst{8}, 
P. Bonifacio\inst{5}, 
C. Allende Prieto\inst{6,7}, 
R. Rebolo\inst{6,7,9}, 
M. Damasso\inst{10},
A. Su\'arez Mascare\~no\inst{6,7}, 
S.~B.~Howell\inst{11},
E. Furlan\inst{12},
S. Cristiani\inst{3},
G. Cupani\inst{3},
P. Di Marcantonio\inst{3}
V. D'Odorico\inst{3,13,4},
C. Lovis\inst{20},
C. J. A. P. Martins\inst{15,16},
D. Milakovic\inst{17},
M. T. Murphy\inst{18},
N.~J. Nunes\inst{14},
F. Pepe\inst{20},
N.~C. Santos\inst{21,22},
T.~M. Schmidt\inst{3,19},
A. Sozzetti\inst{10}}


\institute{Dipartimento di Fisica e Astronomia, Universit \'a degli Studi di Firenze, Via G. Sansone 1, I-50019 Sesto Fiorentino, Italy.
\and 
INAF-Osservatorio Astrofisico di Arcetri, Largo E. Fermi 5, I-50125 Firenze, Italy.
\and
INAF-Osservatorio Astronomico di Trieste, Via G.B. Tiepolo 11, I-34143 Trieste, Italy.
\and
Institute of Fundamental Physics of the Universe, Via Beirut 2, Miramare, Trieste, Italy.
\and
GEPI, Observatoire de Paris, Université PSL, CNRS, 5 Place Jules Janssen, 92190 Meudon, France.
\and
Instituto de Astrof\'{\i}sica de Canarias, V\'{\i}a L\'actea, 38205 La Laguna, Tenerife, Spain.
\and
Universidad de La Laguna, Departamento de Astrof\'{\i}sica,  38206 La Laguna, Tenerife, Spain.
\and
Centro de Astrobiología (CSIC-INTA), Carretera Ajalvir km 4, 28850 Torrejón de Ardoz, Madrid, Spain.
\and
Consejo Superior de Investigaciones Cient\'{\i}ficas, 28006 Madrid, Spain.
\and
INAF - Osservatorio Astrofisico di Torino, Via Osservatorio 20, I-10025 Pino Torinese, Italy.
\and
NASA Ames Research Center, Moffett Field, CA 94035, USA.
\and
NASA Exoplanet Science Institute, Caltech/IPAC, Mail Code 100-22, 1200 E. California Blvd., Pasadena, CA 91125, USA.
\and
Scuola Normale Superiore P.zza dei Cavalieri, 7 I-56126 Pisa.
\and
Instituto de Astrofísica e Ciências do Espaço, Faculdade de Ciências da Universidade de Lisboa,
Campo Grande, PT1749-016 Lisboa, Portugal.
\and
Instituto de Astrof\'isica e Ci\^encias do Espa\c co, CAUP, Universidade do Porto, Rua das Estrelas, 4150-762, Porto, Portugal.
\and
Centro de Astrof\'{\i}sica da Universidade do Porto, Rua das Estrelas, 4150-762 Porto, Portugal.
\and
European Southern Observatory, Karl-Schwarzschild-Str. 2, 85748 Garching bei M\"unchen, Germay.
\and
Centre for Astrophysics and Supercomputing, Swinburne University of Technology, Hawthorn, Victoria 3122, Australia.
\and
Observatoire  Astronomique  de  l’Université  de  Genève,  CheminPegasi 51, Sauverny 1290, Switzerland
\and
Département d’astronomie de l’Université de Genève, Chemin Pegasi 51, 1290 Versoix, Switzerland.
\and
Instituto de Astrofísica e Ciências do Espaço, CAUP, Universidadedo Porto, Rua das Estrelas, 4150-762 Porto, Portugal. 
\and
Departamento de Física e Astronomia, Faculdade de Ciências, Universidade do Porto, Rua Campo Alegre, 4169-007 Porto, Portugal.
\\             
}   

\authorrunning{Aguado, Molaro, et al.\\}
\titlerunning{Binarity properties of HE\,0107$-$5240}

 
  \abstract
   {HE\,0107$-$5240 is a  hyper metal-poor star  with $\rm [Fe/H]=-5.39$,  one of the lowest-metallicity  stars known.  Its stellar atmosphere is enhanced in carbon, with $\rm [C/Fe]=+4.0$, without detectable presence of neutron-capture elements. 
   Therefore, it belongs to the Carbon-Enhanced Metal-Poor (CEMP$-${\it no}) group as the majority of most metal-poor stars known to date.
   Recent studies showed variations in the line-of-sight velocity of HE\,0107$-$5240, suggesting it belongs to  a binary system. CEMP-{\it no} stars are the closest descendants of the very first Pop\,III stars and  binarity  holds important bearings for the poorly known mechanism leading to their formation.}
   {We performed high-resolution observations with the  ESPRESSO spectrograph at the VLT to constrain the kinematical properties of the binary system HE\,0107$-$5240 and to probe the binarity of the sample of 8 most metal-poor stars with $\rm [Fe/H]\le-4.5$.}
   { Radial velocities are obtained by using cross-correlation in the interval 4200$-$4315\,\AA\,, which contains the relatively strong CH band, against a template that could be either a synthetic spectrum or a combined  observed spectrum in an iterative process. A Bayesian method is applied to calculate the orbit by using the ESPRESSO measurements and others from the literature. Chemical analysis has also been performed in HE\,0107$-$5240 emplying spectral synthesis with the {\tt SYNTHE} and {\tt ATLAS} codes. }
   {Observations of HE\,0107$-$5240 spanning more than 3 years show a monotonic decreasing trend in radial velocity at a rate of approximately by 0.5\,\rm m\,s$^{-1}\,$d$^{-1}$. A maximum $v_{rad}$ was reached between March 13th, 2012, and December 8th, 2014. The period is constrained at $P_{\rm orb} = 13009_{-1370}^{+1496}$~d. New more stringent  upper limits have been found for several elements: a)\,[Sr/Fe] and [Ba/Fe] are lower than $-0.76$ and $+0.2$ respectively, confirming the star is a CEMP-{\it no}; b)\,$\rm A(Li)< 0.5$ is well below the plateau at $\rm A(Li)= 1.1$ found in the Lower Red Giant Branch stars, suggesting Li was originally depleted; and c)\,the isotopic ratio $^{12}$C/$^{13}$C is 87$\pm6$  showing very low   $^{13}$C in contrast to what expected from a \textit{spinstar} progenitor.}
   {We confirm that HE\,0107$-$5240 is a binary star with a long period of about 13000\,d ($\sim36$\,years). 
   The carbon isotopic ratio excludes the possibility  that the companion has gone through the AGB phase and transferred mass to the currently observed star.
   The binarity of HE\,0107$-$5240 implies some of the first generations of low-mass stars form in multiple systems and indicates that the low metallicity does not preclude the formation of binaries.  Finally, a solid indication of $v_{ rad}$ variation has been found also in  SMSS\,1605$-$1443.}

               
\maketitle
%

\section{Introduction}\label{intro}
 HE\,0107$-$5240 is a hyper metal-poor\footnotemark{}  star discovered in the context of the Hamburg/ESO survey \citep{chris01b} first reported by \citet{chrisnat}. HE\,0107$-$5240 shows carbon enhancement of  $\rm [C/Fe]=+4.0$  but with no significant n-capture elements enrichment. Therefore, it belongs to the class of the carbon-enhanced metal-poor star called CEMP$-${\it no}, that is $\rm [C/Fe]>1$ and  $\rm [Ba/Fe]<0$ \citep{bee05}. Two years later \citet{fre05} found a very similar star, HE\,1327$-$2326 with $\rm [Fe/H]= -5.6$, which prompted  a  discussion on the frequency of these extremely rare objects \citep[see ][]{cayrel05}. So far, 14 stars with metallicities $\rm [Fe/H]< -4.5$ are known. These stars are the objects with the lowest metallicity measured in primitive Pop\,II stars and provide insight into the nature of the First Stars and on the first chemical production.  The iron abundance of these stars is so low that only a few progenitors, possibly even only one, polluted the gas out of which they were formed. The carbon enhancement has been explained by the explosion of faint supernovae, with energy of $10^{51}$\,erg,  together with fallback and mixing  \citep{ume03}. 
 An alternative explanation is a double source with the lighter elements such as CNO, synthesised by faint SNe and the heavier elements by more conventional core-collapse SNe \citep{boni15}. \footnotetext{Following \citet{bee05} we will refer in this work with ultra metal-poor (UMP), hyper metal-poor (HMP), and mega metal-poor (MMP) to stars with $\rm [Fe/H]<-4.0, -5.0, and -6.0$, respectively.}
 \citet{spi13} noted that the carbon abundance does not follow the iron decrease, but remains rather constant at the lowest metallicity. They suggested the existence of two groups\footnote{Some other authors split the CEMP$-${\it no} into two groups depending on [Fe/H] and [C/Fe] (see \citealt{yoon16, yoon19} for more details).}: the former with high carbon abundance  ($\rm [C/H] \approx -2.0$), and a latter with lower  carbon abundance ($\rm [C/H] \approx -3.5$).  The former group also shows enhancement  of n-capture elements with $\rm [Ba/Fe]>1$, which is absent in the latter group. A different origin for carbon has been suggested in the two groups. In the \textit{High-C} group,  both carbon and the  n-capture elements come from an asymptotic giant branch (AGB) companion, while the origin of carbon in the \textit{Low-C} band is less certain, but should be pristine and should be coming from the SNe that polluted  the gas from which the star is formed. This scenario implies that the \textit{High-C} group consists of binaries while the \textit{Low-C} group could be single or binary stars, but without a companion which experienced an AGB phase. This hypothesis has been tested with radial velocity studies in a small sample of metal-poor stars belonging to the two groups \citep{han16I,han16III}. Additional CEMP stars with [Fe/H]$<$ -4.5 have been recently discovered  revealing that all the CEMP stars with $\rm [Fe/H] <-4.5$ belong to the  CEMP-{\it no} group \citep{boni18,agu18I,nordlander2019MNRAS.488L.109N, jon20}, and  therefore  are the preferred formation channel for the first low mass star formation. 
 
 HE\,0107$-$5240 was employed by several groups to constrain the mass of the very first stars \citep[see e.g.][]{boni03,Schneider03,ume03}. The existence of stars like  HE\,0107$-$5240 shows  that stars with masses of the order of the sun or smaller do form at the lowest metallicities.
    \citet{bro03} argued that  the significant values of carbon and oxygen could have cooled the gas to permit low-mass star formation. However, the discovery of carbon normal stars such as  SDSS\,J1029+1729  \citep{caff11} and of Pristine221.8781+9.7844 \citep{sta18} shows that not all the extremely metal-poor stars are carbon enhanced. More complex mechanisms for gas cooling such as dust cooling or turbulent fragmentation are required \citep{sche12,gre12}.

The chemical study of HE\,0107$-$5240 is remarkably challenging. Due to the extreme low metallicity, only nine elemental abundances (C, N, O, Na, Mg, Ca, Ti, Fe, and Ni) have been measured so far \citep{chrisnat, chris04, coll06}. Significant upper limits were only possible for two n-capture elements, $\rm [Sr/Fe]< -0.52$ and $\rm [Ba/Fe]< 0.82$ \citep{chris04}. Understanding its kinematics was not simpler: more than 15 years after the discovery of HE\,0107$-$5240, the first indication of binarity was found by \citet{arentsen19}, which opened the possibility of a mass transfer from an AGB companion.  This unexpected binarity behaviour was later confirmed by \citet{boni20}. Historically, it is interesting to note that the prototype of the CEMP-{\it no} star CS\,22957$-$027, discovered almost simultaneously by \citealt{norris1997ApJ...489L.169N,bonifacio1998A&A...332..672B},  has also been shown to be a binary star with a period of about  $\rm P=3125$\,d by \citet{preston2001AJ....122.1545P}. 

In this work, we continue the radial velocities measurements with the aim to detail the binary properties of this ancient star. At the same time, we monitor the radial velocities, $v_{rad}$, for other CEMP-{\it no} stars with $\rm [Fe/H]< -4.5$ to assess the fraction of binaries in this group.


\begin{table}
\begin{center}
    
\caption{\label{table:param} Targets observed in our ESPRESSO GTO program.}
\scriptsize
\hspace*{0cm}\resizebox{1.\linewidth}{!}{
\begin{tabular}{lcccccc}
 \hline
 STAR&{\it Gaia} G&  \teff & \logg & [Fe/H] & [C/Fe]&RUWE$^{a}$\\
   &mag& K & cgs &  & &\\
 \hline
  \hline
HE\,0107$-$5240$^{1}$   &14.9&5100 &2.2   &$-5.39$ &4.00 &1.01 \\
SDSS\,J0023+0307$^{2,3}$  &17.6&6140 &4.8   &$<-5.50$&>3.31&0.95 \\ 
HE\,0233$-$0343$^{4}$  &15.3 &6100 &3.4   &$-4.68$ &3.32 &1.02 \\
SMSS\,0313$-$6708$^{5}$ &14.5&5125 &2.3   &$<-7.10$&>5.39&1.02 \\
HE\,0557$-$4840$^{6}$   &15.2&4900 &2.2   &$-4.75$ &1.66 &1.07 \\
SDSS\,J1313$-$0019$^{7,8}$&15.2&5170 &2.6   &$-5.00$ &3.00 &0.99 \\
HE\,1327$-$2326$^{9}$  &13.4 &6180 &3.7   &$-5.60$ &4.26 &0.98   \\
SMSS\,1605$-$1443$^{10}$ &15.4&4850 &2.0   &$-6.21$ &3.89 &1.06  \\
\hline
\end{tabular}}
$^{a}$The Renormalised Unit Weight Error could be used to detect variability (see text).
\tablefoot{{(1) \citet{chris02}; (2) \citet{agu18II}; (3) \citet{agu19a}; (4) \citet{han14}; (5) \citet{kel14}; (6) \citet{nor07}; (7) \citet{alle15}; (8) \citet{fre15}; (9) \citet{fre05}; (10) \citet{nordlander2019MNRAS.488L.109N}.}}
\end{center}
\end{table}

\begin{figure}
\begin{center}
{\includegraphics[width=90 mm, angle=0]{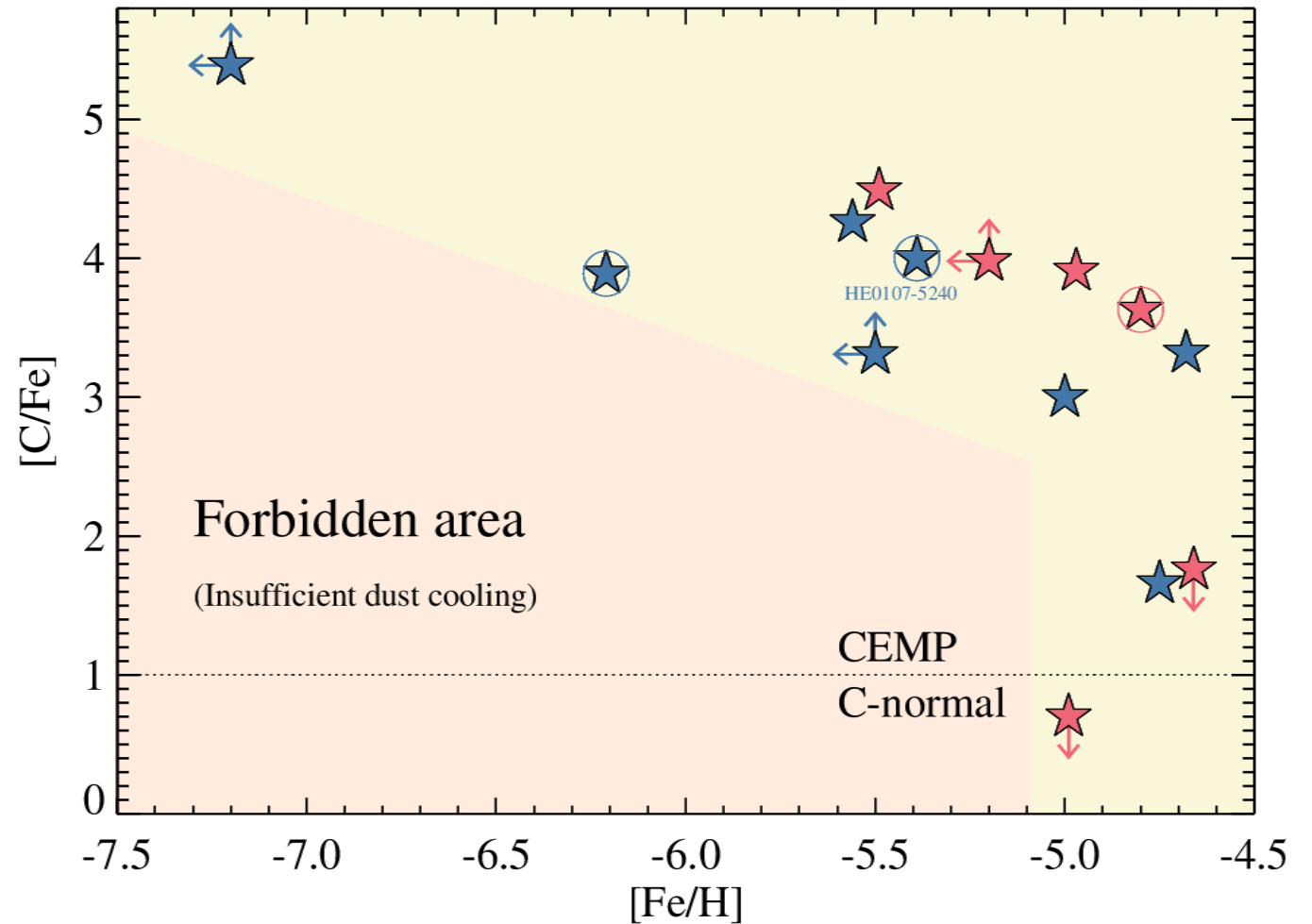}}
\end{center}
\caption{[C/Fe]$-$[Fe/H] plane for all the stars observed with ESPRESSO (blue symbols), and for other metal-poor stars (red symbols). Stars with confirmed (HE\,0107$-$5240 and SDSS\,J0929+0238) and indicated (SMSS\,16054$-$1443) binarity are marked with a circle. According to \citet{chi17}, the region where no low-mass star formation is permitted is also red-filled.}
\label{fig:carbon}
\end{figure}

\begin{figure*}
\begin{center}
{\includegraphics[width=190 mm, angle=0]{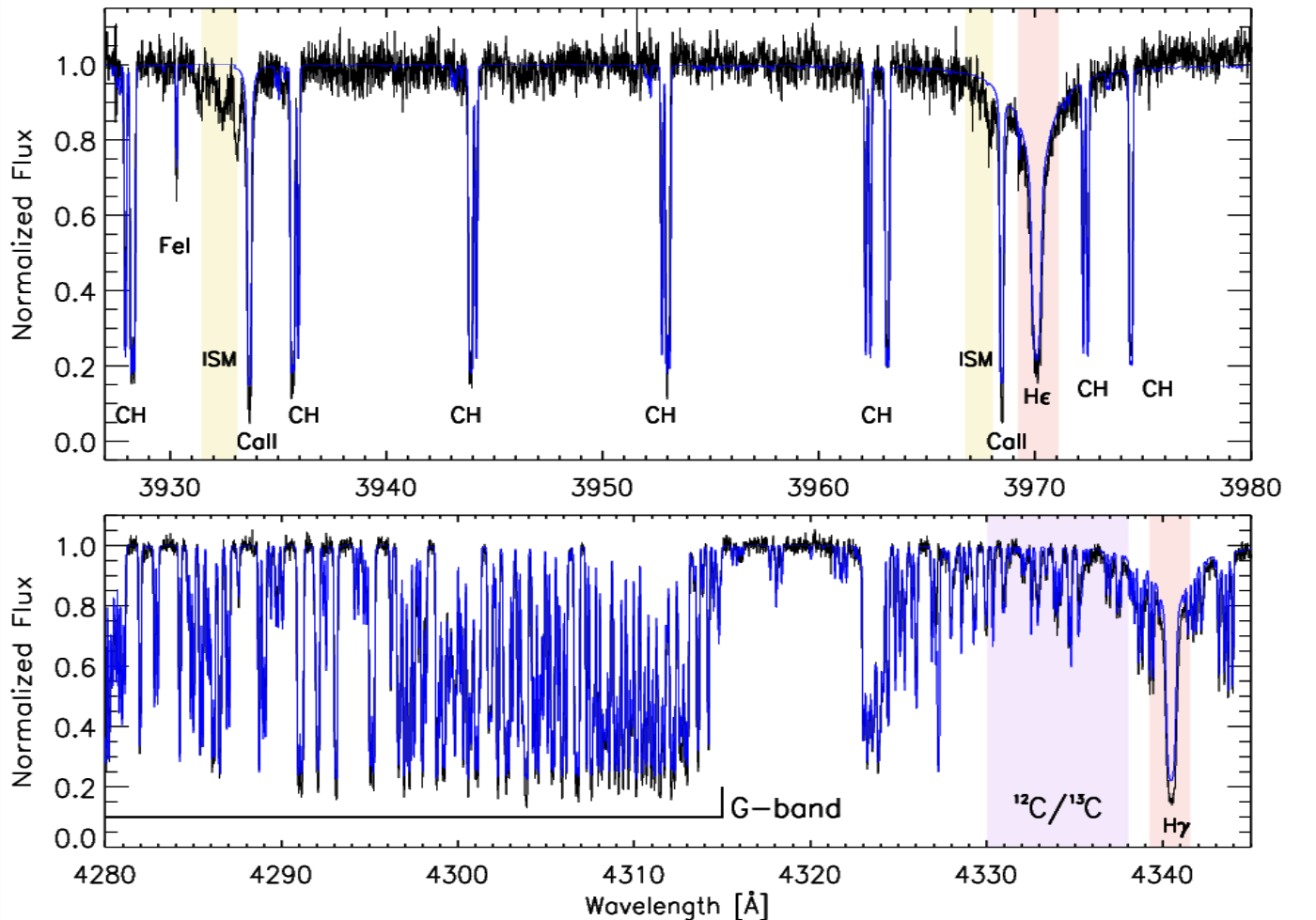}}
\end{center}
\caption{A narrow region of the co-added ESPRESSO spectrum of HE\,0107$-$5240 (black  lines), around the Ca H\&K area (upper panel)  and  the G-band (lower panel)  together  with a SYNTHE model computed to match the HE\,0107$-$5240 stellar parameters (blue line). Asborptions corresponding to the Balmer series and interstellar \ion{Ca}{ii} are marked in red and yellow respectively. The region from which we measure the $^{12}$C/$^{13}$C ratio is marked in viola. Other metallic absorptions and the G-band, are also labelled.}
\label{comp}
\end{figure*}

\section{ESPRESSO Observations and Data Reduction}\label{sec:obs}
The  Echelle Spectrograph for Rocky Exoplanets and Stable Spectroscopic Observations (ESPRESSO), is a stable fibre-fed \textit{échelle} spectrograph mounted at the incoherent focus of the Very Large Telescope (VLT) at Paranal Observatory, Chile, designed to measure velocities with precision as high as 10\,$\rm cm\,s^{-1}$ \citep[ESPRESSO, ][]{pepe21}. In  2018 we started an ESO$-$GTO program, P.I. P. Molaro,  to monitor radial velocities of the most metal-poor stars to probe their binarity.
The spectrograph has two fibers with a core diameter of $140\,\mu m$ corresponding to a 1\farcs0 aperture in the sky.  While Fibre A was centered onto the object, Fibre B was in the sky. The binning of the CCD was initially  $2\times1$ pixel and, when it became available, $4\times2$, always with a slow readout mode. Observations were done in single UT configuration mode with exposure times between t$_{exp}=1800$ and t$_{exp}=3600$\,s, the maximum allowed by service mode operation. A total of $~50$\,h (time on target) have been observed during this -still ongoing- program. The observing conditions were restricted as follows: airmass$\leq 1.5$, seeing $\leq 1\farcs0$, water vapour $\leq 30$\,mm and minimum lunar distance $=30^{\circ}$.
Data reduction was performed by the automatic ESPRESSO pipeline including sky substraction, bias and flat-fielding correction. The wavelength calibration is the one which combines the  ThAr lamp  with a Fabry-P\'erot Etalon \citep[][]{pepe13}. The signal-to-noise ratio of individual exposures close to the G-band ($\lambda \lambda \sim 4300$~\AA) spanned the range $5 \leq {\rm SNR} \leq 20$, depending on the observing time and the magnitude of the target. 
We retrieved an additional exposure taken with HARPS at the ESO 3.6\,m telescope of HE\,0107$-$5240. The observation was made in Technical Time in the high-resolution mode ``HAM''. Observing time was 3600\,s and $\rm SNR\sim2$ at 4300\,\AA~with a seeing of 0.75".

While our primary target is HE\,0107$-$5240, with 17 visits, the program also included ESPRESSO observations for seven more CEMP-{\it no} stars with $\rm [Fe/H]<-4.5$. In table \ref{table:param} all the observed targets and their main stellar parameters are summarised. The situation of these primitive stars in the [C/Fe]$-$[Fe/H] plane is shown in Fig. \ref{fig:carbon}, together with other metal-poor stars that are not included in our program for various reasons that will be discussed in Section \ref{sec:others}.

\section{Radial Velocity Determination}\label{sec:vrad}
The ESPRESSO radial velocity precision allows us to detect binary systems even if they show long periods (up to several decades) and relatively small radial velocity changes in a period of time of a few years.  
\begin{table*}
\begin{center}
    
\caption{\label{table:rvs} Radial velocities measurements from ESPRESSO.}
\scriptsize
\hspace*{0cm}\resizebox{1.\linewidth}{!}{
\begin{tabular}{lccccccc}
 \hline
 STAR & $v_{rad}$ & error & MJD\tablefootmark{a} & MODE&t$_{exp}$&CCF&Comment\tablefootmark{b}\\
   & (km$\,$s$^{-1}$) & (km$\,$s$^{-1}$) & $-50000$ & &s & \\
 \hline
  \hline
HE\,0107$-$5240   &48.049    &  0.034 & 8364.107 & HR21 &3300& CH (G$-$band) & \\
HE\,0107$-$5240 &48.083&0.104&8473.127& - &3600&CH (G$-$band) &HARPS\\
HE\,0107$-$5240   & 47.833   &  0.010 & 8698.313 & HR21 &3400& CH (G$-$band) & \\
HE\,0107$-$5240   & 47.771   &  0.010 & 8721.263 & HR21 &3400& CH (G$-$band) &\\
HE\,0107$-$5240   & 47.786   &  0.009 & 8730.206 & HR21 &3400& CH (G$-$band) &\\
HE\,0107$-$5240   & 47.818   &  0.012 & 8741.250 & HR21 &3400& CH (G$-$band) &\\
HE\,0107$-$5240   & 47.846   &  0.012 & 8759.096 & HR21 &3400& CH (G$-$band) &\\
HE\,0107$-$5240   & 47.697   &  0.012 & 9190.096 & HR21 &3417& CH (G$-$band) &\\
HE\,0107$-$5240   & 47.625   &  0.012 & 9237.032 & HR42 &3417& CH (G$-$band) &\\
HE\,0107$-$5240   & 47.562   &  0.018 & 9427.332 & HR21 &3600& CH (G$-$band) &\\
HE\,0107$-$5240   & 47.533   &  0.012 & 9435.176 & HR42 &3600& CH (G$-$band) & \\
HE\,0107$-$5240   & 47.542   &  0.010 & 9435.219 & HR42 &3600& CH (G$-$band) & \\
HE\,0107$-$5240   & 47.508   &  0.013 & 9489.238 & HR21 &3000& CH (G$-$band) & \\
HE\,0107$-$5240   & 47.505   &  0.009 & 9618.061 & HR42 &3310& CH (G$-$band) & \\
HE\,0107$-$5240   & 47.497   &  0.011 & 9529.202 & HR42 &3200& CH (G$-$band) &\\
HE\,0107$-$5240   & 47.509   &  0.008 & 9545.123 & HR42 &3200& CH (G$-$band) &\\
HE\,0107$-$5240   & 47.436   &  0.013 & 9606.042 & HR42 &3600& CH (G$-$band) &\\
SDSS\,J0023+0307  & -195.540 &  0.107 & 8761.245 & HR21 &3400&\ion{Mg}{i}\,b& Synthetic template\,\,\,\,\,\,\,\, \\ 
SDSS\,J0023+0307  & -195.035 &  0.171 & 8805.034 & HR21 &3400&\ion{Mg}{i}\,b& Synthetic template\,\,\,\,\,\,\,\, \\
SDSS\,J0023+0307  & -195.761 &  0.102 & 8835.044 & HR21 &3400&\ion{Mg}{i}\,b& Synthetic template\,\,\,\,\,\,\,\,\\
HE\,0233$-$0343   & 52.058   &  0.027 & 8699.288 & HR21 &3400& CH (G$-$band)    & \\
HE\,0233$-$0343   & 52.127   &  0.021 & 8760.318 & HR21 &3400& CH (G$-$band)    & \\
HE\,0233$-$0343   & 52.169   &  0.024 & 8783.112 & HR21 &3400& CH (G$-$band)    & \\
HE\,0233$-$0343   & 51.074   &  0.024 & 8813.243 & HR21 &3400& CH (G$-$band)    & \\
SMSS\,0313$-$6708 & 298.643  &  0.043 & 8701.281 & HR21 &3400& CH (G$-$band)    & \\
SMSS\,0313$-$6708 & 298.677  &  0.042 & 8732.245 & HR21 &3400& CH (G$-$band)    & \\
SMSS\,0313$-$6708 & 298.509  &  0.033 & 8732.343 & HR21 &3400& CH (G$-$band)    & \\
SMSS\,0313$-$6708 & 298.710  &  0.043 & 8740.153 & HR21 &3400& CH (G$-$band)    & \\
SMSS\,0313$-$6708 & 298.581  &  0.039 & 8740.196 & HR21 &3400& CH (G$-$band)    & \\
SMSS\,0313$-$6708 & 298.632  &  0.046 & 8742.119 & HR21 &3400& CH (G$-$band)    & \\
SMSS\,0313$-$6708 & 298.578  &  0.042 & 8743.268 & HR21 &3400& CH (G$-$band)    & \\
SMSS\,0313$-$6708 & 298.531  &  0.047 & 8743.335 & HR21 &3400& CH (G$-$band)    & \\
SMSS\,0313$-$6708 & 298.536  &  0.050 & 8780.153 & HR21 &3417& CH (G$-$band)    & \\
SMSS\,0313$-$6708 & 298.650  &  0.036 & 9190.052 & HR42 &3417& CH (G$-$band)    & \\
SMSS\,0313$-$6708 & 298.603  &  0.038 & 9237.078 & HR42 &3417& CH (G$-$band)    & \\
SMSS\,0313$-$6708 & 298.530  &  0.027 & 9250.041 & HR42 &3417& CH (G$-$band)    & \\
SMSS\,0313$-$6708 & 298.486  &  0.039 & 9264.017 & HR42 &3000& CH (G$-$band)    & \\
HE\,0557$-$4840   & 212.209  &  0.028 & 8761.333 & HR21 &3400&\ion{Mg}{i}\,b     &\\
HE\,0557$-$4840   & 212.172  &  0.026 & 8762.327 & HR21 &3400&\ion{Mg}{i}\,b     &\\
HE\,0557$-$4840   & 212.199  &  0.027 & 8834.084 & HR21 &3400&\ion{Mg}{i}\,b     &\\
HE\,0557$-$4840   & 212.198  &  0.023 & 8864.083 & HR21 &3400&\ion{Mg}{i}\,b     &\\
SDSS\,J1313$-$0019& 273.984  &  0.054 & 9649.249 & HR42 &3444& CH (G$-$band)    & Synthetic template\\
HE\,1327$-$2326   & 63.801   &  0.040 & 8615.137 & HR21 &1800& CH (G$-$band)    &\\
HE\,1327$-$2326   & 63.636   &  0.037 & 8688.022 & HR21 &2700& CH (G$-$band)    &\\
HE\,1327$-$2326   & 63.709   &  0.041 & 8688.055 & HR21 &2700& CH (G$-$band)    &\\
HE\,1327$-$2326   & 63.655   &  0.027 & 8695.996 & HR21 &1800& CH (G$-$band)    &\\
HE\,1327$-$2326   & 63.639   &  0.037 & 8697.988 & HR21 &2400& CH (G$-$band)    &\\
HE\,1327$-$2326   & 63.697   &  0.041 & 8841.331 & HR21 &1800& CH (G$-$band)    &\\
HE\,1327$-$2326   & 63.601   &  0.036 & 9429.054 & HR42 &3000& CH (G$-$band)    &\\
HE\,1327$-$2326   & 63.700   &  0.016 & 9663.072 & HR42 &2844& CH (G$-$band)    &\\
SMSS\,1605$-$1443 &-226.059  &  0.037 & 9676.277 & HR42 &3150& CH (G$-$band)    &\\
SMSS\,1605$-$1443 &-226.118  &  0.036 & 9726.189 & HR42 &3150& CH (G$-$band)   &  \\
SMSS\,1605$-$1443 &-226.196  &  0.037 & 9727.060 & HR42 &3150& CH (G$-$band)    &\\
SMSS\,1605$-$1443 &-226.214  &  0.030 & 9761.059 & HR42 &3150& CH (G$-$band)    &\\
SMSS\,1605$-$1443 &-226.278  &  0.033 & 9792.055 & HR42 &3150& CH (G$-$band)    &\\
\hline
\end{tabular}}
\tablefoot{\tablefoottext{a}{Modified Julian date at the start of observation.}\tablefoottext{b}{Measurements obtained from a synthetic template are indicated (See text for details).}}
\end{center}
\end{table*}

The  measurements of the stellar radial velocities are performed in two steps. \begin{itemize}
    \item First, a synthetic model with the {\tt SYNTHE}  code \citep{kur05, sbo05} is computed by assuming stellar parameters and abundances from the literature values \citep{chris04}. Then we smoothed and resampled the synthetic spectra to the ESPRESSO (and HARPS) resolving power of $\rm R\sim$140,000 (and $\rm R\sim$115,000) and to the same lambda step. Subsequently,  we normalised the observed spectra and interpolated both data and the template to the same wavelength. A cross-correlation function (CCF) was performed in the Fourier Space by using the algorithm by \citet{tonry79} which contained the {\tt noao.rv} package within {\tt IRAF}\footnote{{\tt IRAF} is distributed by the National Optical Astronomy Observatory which is operated by the Association of Universities for Research in Astronomy (AURA) under a cooperative agreement with the National Science Foundation.} \citep{tod93} environment over the 4200$-$4315\,\AA\, interval, where the CH absorptions are detectable in individual exposures. 
    \item Second,  we corrected each exposure by the shift derived from the CCF and combined all the  spectra in the rest of frame with a   3$\sigma$-clip algorithm to compute a high SNR ``natural'' template.  An example of this coadded spectrum of HE\,0107$-$5240 is shown in Fig. \ref{comp}. Similarly to the previous step, we normalised the combined spectrum and calculated a new CCF for each individual exposure obtaining  our final $v_{rad}$ values. 
\end{itemize}

The radial velocities  given by the CCF are summarised in Table \ref{table:rvs}. Focusing in HE\,0107$-$5240, the average error we have from ESPRESSO data is 13\,m$\,$\rm s$^{-1}$, 8 times lower than the one from HARPS, 104\,m$\,$\rm s$^{-1}$.
We also notice the  radial velocity of the \#1 ESPRESSO observation taken on 3rd September 2018 is in common with \citet{boni20} and the two measurements are  consistent at 1.5$\sigma$. This small difference is mostly attributed to the fact that  the CCF of \citet{boni20} is performed over a different  range, 4000$-$4498\,\AA, including other lines which show  some velocity offset with respect to the CH molecular absorption. In the next section we present a complete kinematical analysis based in the results we obtained for HE\,0107$-$5240 from ESPRESSO data and from other sources in the literature.

\section{Binarity of   HE\,0107$-$5240}\label{sec:anal}

The radial velocities of HE\,0107$-$5240 show a monotonically decreasing trend  and therefore we confirm that this is a binary star with a long period. The overall rate of change in this period is of  $\sim$ 0.5\,\rm m\,s$^{-1}\,$d$^{-1}$, which  is consistent with  previous results by \citet{arentsen19, boni20}. 

\begin{figure*}
\begin{center}
{\includegraphics[width=170 mm, trim={ .25cm 0.cm 0.3cm 0.cm},clip]{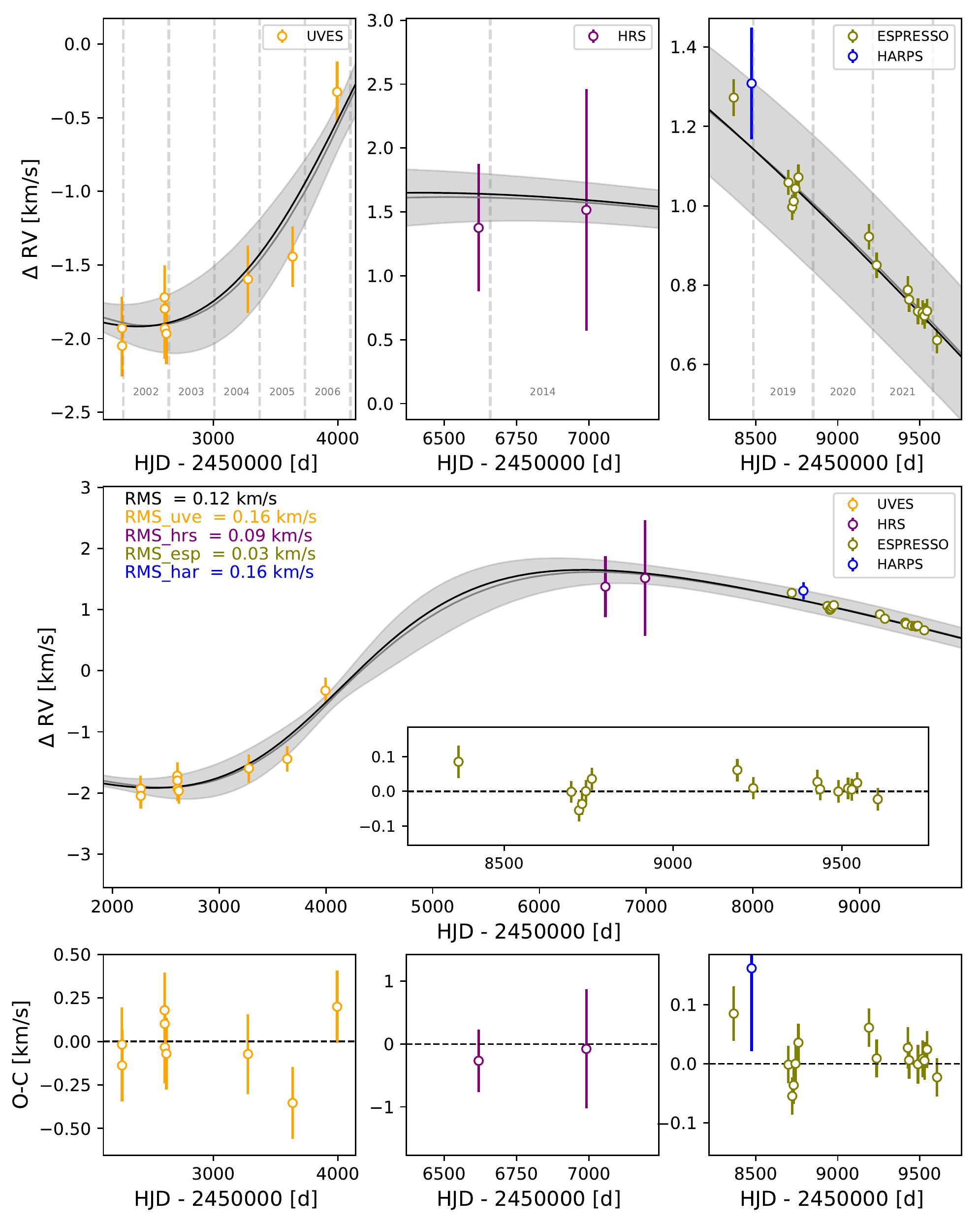}}
\end{center}
\caption{Radial velocity (RV) points of HE\,0107$-$5240 versus heliocentric Julian date (HJD), together with the best RV model. The inner panel within the middle panel shows the ESPRESSO RV points only after removing the best RV model. Within the middle panel we provide the RMS of the residuals of all RVs and those from each spectrograph given in the bottom panels.}
\label{vrad}
\end{figure*}

To constrain the kinematical properties of the binary star HE\,0107$-$5240 together with the ones derived in this work we take advantage of previous $v_{rad}$ measurements from \citet[UVES at VLT]{chris04}, \citet[HRS at SALT]{arentsen19} and \citet[ESPRESSO at VLT]{boni20}. The $v_{rad}$ data are binned in 1$-$d bins to limit the stellar jitter. This only affects the UVES data reducing the number of points from 33 to 17 (See Fig. \ref{vrad}, upper-left panel). The HRS points with $v_{rad}$ errors larger than 2.0\,km$\,$s$^{-1}$ have been discarded from Fig. \ref{vrad} (upper-middle panel). All the RVs are modelled using  the {\sc radvel}~\citep{fulton2018radvel}\footnote{ {\sc radvel} is a Python package for modeling of radial velocity time series data, available in https://radvel.readthedocs.io/en/latest/index.html }  including a Keplerian motion of the star plus an RV offset (the $\gamma$ of the $v_{rad}$ curve), and a jitter parameter for each instrument  within a likelihood scheme implemented in python using {\sc celerite}~\citep{celerite}\footnote{{\sc celerite} is a library for fast and scalable Gaussian Process (GP) Regression in one dimension, available in https://celerite.readthedocs.io/en/stable/ } . Unfortunately, there are no contemporaneous observations of different instruments, except for the single HARPS RV point, that would allow us to derive any possible RV offset between different instruments, but the jitter may be absorbing part of this possible offset with an uncertainty which propagates onto the uncertainties on orbital period, semi-amplitude velocity and eccentricity. The Keplerian orbit is described as:
\begin{equation}
         v_{rad} = \gamma + k_{2} (\cos(\nu_{a} + \omega) + e \cos(\omega))
\end{equation}
where the true anomaly $\nu_{a}$ is related to the solution of the Kepler equation that depends on the orbital period, $P_{2}$, the orbital semi-amplitude velocity, $k_2$, the argument of periastron, $\omega$, and the eccentricity of the orbit, $e$. 
The {\sc radvel} allows to infer both the time of periastron, $T_{\rm 2,peri}$, and the time at inferior conjunction of the star, $T_{\rm 2,conj}$. 
Following \citet{fulton2018radvel}, we chose to perform fitting and posterior sampling using the $\ln P_2$, $T_{\rm 2,conj}$, $\sqrt{e}\cos \omega$, $\sqrt{e}\sin \omega$, $\ln k_{2}$ basis which helps to speed the convergence. We chose $\ln k_{2}$ to avoid favouring large $k_{2}$ and $\ln P_2$ because $P_2$ may be long if compared with the observational baseline.
The $v_{rad}$ errors in the Fig.~\ref{vrad} include the jitter of each instrument and the $v_{rad}$ offset has been subtracted. In addition, we adopted uniform priors for all parameters (see Table~\ref{tab:par}).

\begin{figure*}
\begin{center}
{\includegraphics[width=180mm, trim={ 0.5cm 0.cm 1.cm 0.cm},clip]{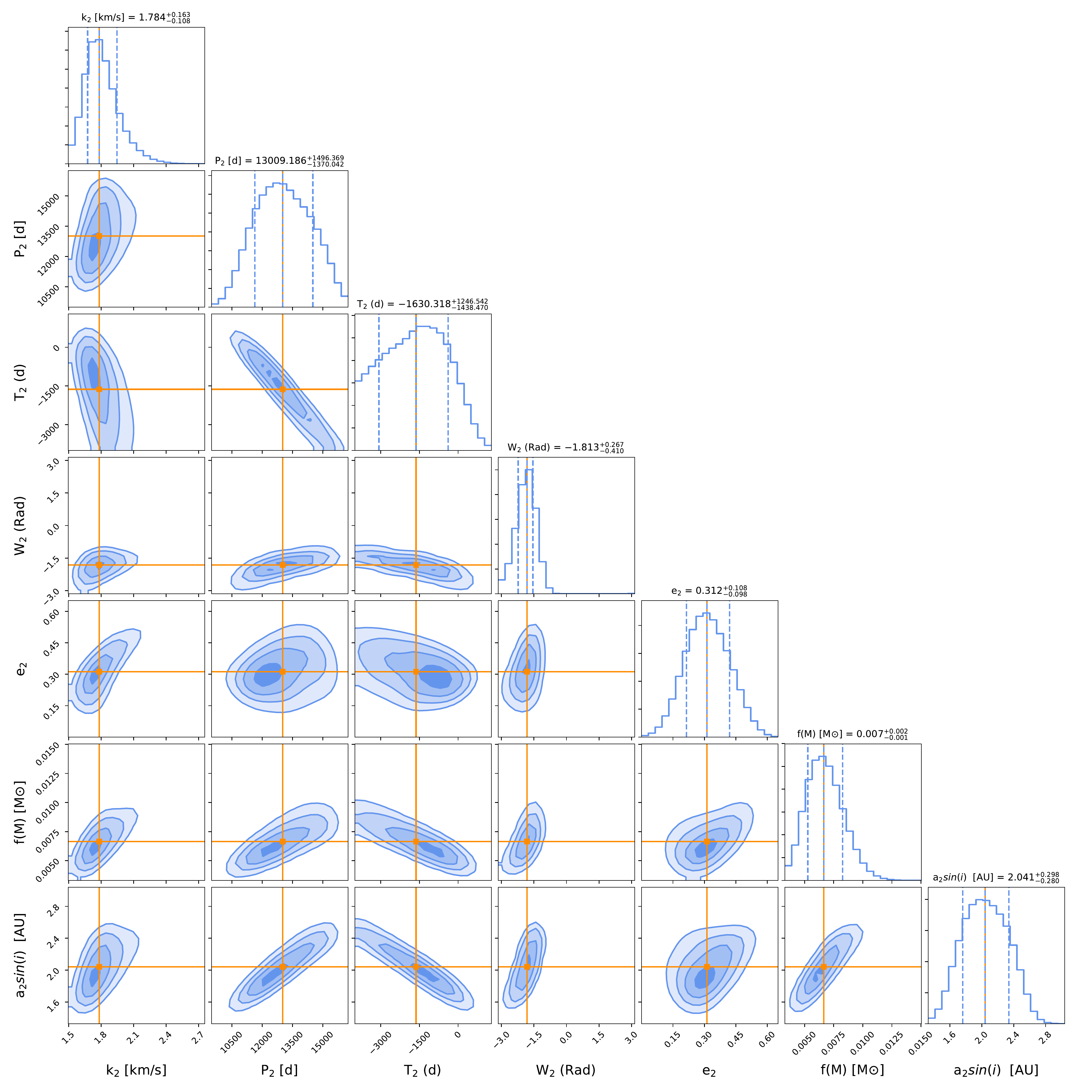}}
\end{center}
\caption{Posterior distributions of the star's orbital parameters of the RV model and two additional parameters, the binary mass function, $f({\rm M})$, and the star's orbital distance to the center of mass, $a_2\sin i$, inferred from the orbital parameters.}
\label{corner}
\end{figure*}

\begin{table}
\centering
	\caption{Orbit parameters of HE\,0107$-$5240 from the MCMC analysis. }
	\label{tab:par}
	\begin{tabular}{lcr} 
		\hline
		Parameter & MCMC Priors & Results \\
		\hline
				&  \multicolumn{2}{c}{Keplerian orbit} \\
		\hline
		$k_2$ [km\,s$^{-1}$]  & $\mathcal{LU}$ (0.1, 5)       & 1.78$^{+0.16}_{-0.11}$    \\ 
		$P_2$ [d]           & $\mathcal{LU}$ (1000, 75000)  & 13009$^{+1496}_{-1370}$    \\
		$T_2-2450000$ [d]   & $\mathcal{U}$ (-4000, 3500)   & -1630$^{+1246}_{-1438}$    \\
		$\rm \omega$ [rad]  & $\mathcal{U}$ (-$\pi$, $\pi$) & -1.81$^{+0.27}_{-0.41}$    \\
		$e$                 & $\mathcal{U}$ (0, 1)          & 0.31$^{+0.11}_{-0.10}$ \\
		\hline
		& \multicolumn{2}{c}{Other terms} \\
		\hline
		jitter$_{\rm UVES}$     [km\,s$^{-1}$] & $\mathcal{LU}$ (0.01, 5.0) & 0.21$^{+0.06}_{-0.05}$ \\
		jitter$_{\rm HRS}$      [km\,s$^{-1}$] & $\mathcal{LU}$ (0.01, 5.0) & 0.09$^{+0.29}_{-0.07}$ \\
		jitter$_{\rm ESPRESSO}$ [km\,s$^{-1}$] & $\mathcal{LU}$ (0.01, 5.0) & 0.03$^{+0.01}_{-0.01}$ \\
		jitter$_{\rm HARPS}$    [km\,s$^{-1}$] & $\mathcal{LU}$ (0.01, 5.0) & 0.09$^{+0.24}_{-0.07}$ \\
		$\gamma-46.5$           [km\,s$^{-1}$] & $\mathcal{U}$ (-3.0, 3.0)  & 0.27$^{+0.14}_{-0.19}$ \\
		\hline
	\end{tabular}
	\begin{minipage}{\columnwidth} 
\end{minipage}	
\end{table}

 To sample the posterior distributions and obtain the Bayesian evidence of the model (that is, marginal likelihood, ln Z), we relied on nested sampling using {\sc dynesty} \citep{speagle20dynesty}\footnote{ {\sc dynesty} is a Pure Python, MIT-licensed Dynamic Nested Sampling package for estimating Bayesian posteriors and evidences, available in https://dynesty.readthedocs.io/en/stable/ }. We initialized a number of live points equal to $N^3(N + 1)$, to efficiently sample the parameter space, with $N = 10$ being the number of free parameters.
The resulting posterior distributions are displayed in Fig. \ref{corner} together with the orbital parameters of the star. In the middle panel of Fig. \ref{vrad} we also show in grey a subsample of 300 models randomly selected from the final selection of $\sim$37,696 Bayesian samples from the posterior distributions displayed in Fig. \ref{corner}. 
On the other hand, the binary mass function can be computed from the masses of the binary component \citep[e.g.][]{TvdH06}, i.e. that of the seen star, $M_2$, and that of the unseen binary companion, $M_1$, and the orbital parameters as follows:
\begin{equation}
         f({\rm M}) = (M^{3}_{1} \sin^{3} i )/ (M_{1} + M_{2})^2
\end{equation}
\begin{equation}
        f({\rm M}) = [(P_{\rm orb} k_{2}^{3} )/ (2\pi G)] (1-e^{2})^{3/2}
\end{equation}

With the  Bayesian samples we computed the posterior distribution of the
binary mass function providing the following result:
\begin{equation}
   f({\rm M}) = 0.007_{-0.001}^{+0.002} \, [M_{\odot}] = 7.0_{-1.4}^{+1.7} \, [M_{\rm Jup}]
\end{equation}

Depending on the ratio of masses between the seen and unseen objects of the binary, usually the binary
mass function provides a lower limit to the mass of the unseen companion. In general, the minimum mass of
this companion can be approximately estimated as the maximum value between $f$ and $f^{1/3} M_2^{2/3}$, thus giving $M_1 [M_{\odot}] > 0.16$, when adopting a mass of $M_2[M_{\odot}]=0.8$ for the seen star HE\,0107$-$5240.

The minimum orbital distance (or semi-major axis) of the star with respect to the center of mass of the binary system can be computed as :

\begin{equation}
        a_{2}\sin{i} = [(P_{\rm orb} k_{2})/(2\pi)] (1-e^{2})^{1/2} 
\end{equation}

The orbital period of the binary is $35.6\pm4.1$~yr. Using the Bayesian samples we also calculated the posterior distribution of the minimum orbital distance (or semi-major axis of the star's orbit) at $a_{2}\sin{i} = 2.04\pm0.30$\,AU (see Fig.~\ref{corner}). We note, as a comparison, Jupiter's orbit in the solar system shows a semimajor axis of about 5.2\,AU and an orbital period of 11.9 yr.

\section{A hint for a secondary motion?}
The dispersion of the ESPRESSO radial velocities after subtracting the best fit is 35\,m\,s$^{-1}$, which is in line with the derived velocity jitter (Table~\ref{tab:par}). However, this is about 3 times greater than the mean error bar of the ESPRESSO radial velocity measurements, suggesting the presence of additional signal of relatively smaller amplitude. 
Careful inspection of the ESPRESSO velocity residuals (see bottom panel of Fig.~\ref{vrad}) indicates variability time scales of the order of tens to hundreds of days.
We analyzed the residuals after the fit using the Generalized Lomb Scargle periodogram \citep[GLS,][]{Zechmeister09}. The GLS is a an algorithm for detecting periodicities in unevenly sampled time-series, equivalent to least-squares fitting of sine waves, that generalizes the Lomb-Scargle periodogram   \citep{scargle82} by including a floating zero-point and weights to the individual datapoints. We computed the GLS between 2 and 1000\,d and found no significant peak with a false-alarm probability smaller than 10\%
This is not surprising given the small number of ESPRESSO data.

One possible scenario to account for the variability of the ESPRESSO velocity residuals is the existence of pulsations in HE\,0107$-$5240. However, oscillations have their characteristic frequency at shorter time scales. Theory predicts that the longest period for first overtone pulsators increases as the metal content decreases, being of about 9 days at $Z = 0.0004$ and of about 6 days at $Z = 0.004$ \citep{marconi10}. \cite{creevey19} reported the first detection of sun-like oscillations in the moderately metal poor giant star HD\,122563 (Fe/H] = $-2.80 \pm 0.15$ \citep[][]{jofre14} using hundreds of radial velocity measurements obtained with a mean cadence of one observation per day. They found a characteristic frequency at 3.07\,$\pm$\,0.05 $\mu$Hz (3.77\,$\pm$\,0.06 d) for the oscillations. Interestingly, they also found an additional signal in the form of a long-term trend with a velocity amplitude and time scale similar to those of HE\,0107$-$5240. In HD\,122563, the velocity amplitude and period of the trend are about 120 m\,s$^{-1}$ and 300 d, respectively. \cite{creevey19} favored stellar accretion, activity, and rotation against oscillations to explain the trend because the time scale of the long-term variation is far from the expected intrinsic pulsation periods.

A second scenario is stellar activity that may produce variations in the ESPRESSO radial velocities modulated by the rotation of the star. \cite{ceillier17} found that $\approx$2\,\%~of the giant stars are actually fast rotators with rotation periods of less than 170 d.  Adopting the mass of 0.8 $\pm$ 0.1\,M$_\odot$ and the surface gravity of log\,$g [{\rm cm\,s}^{-2}]$ = 2.2 $\pm$ 0.1 \citep{chris04}, the estimated stellar radius of HE\,0107$-$5240 is determined at 11.8\,$^{+2.2}_{-2.0}$ R$_\odot$, which combined with periods of 50--200\,d (time-scale of the radial velocity variability) would yield equatorial rotational velocities of 2--13 km\,s$^{-1}$ when the star is seen equator-on. This rotational velocity is measurable with high-resolution spectrographs like ESPRESSO, but we do not detect any significant rotational broadening in the spectra ($v$\,sini\,$i \le 2$ km\,s$^{-1}$). This implies that either HE\,0107$-$5240's rotation axis is likely inclined, or periods of ten to a few hundred days are too short for them to be ascribed to the star's rotation. The Transiting Exoplanet Survey Satellite ({\sl TESS}, \citealt{ricker15}) observed HE\,0107$-$5240 in sectors 2 (2018 Aug 23 -- 2018 Sep 19) and 29 (2020 Aug 26 -- 2020 Sep 21). The exposure times were 1426\,s for the former and 475\,s for the latter sector. Each {\sl TESS} sector covers about 25\,d (slightly shorter than the time scales of the spectroscopic variability) and the two sectors were observed two years apart. We downloaded the target pixel file products from the Mikulski Archive for Space Telescopes and processed them with the {\em lightkurve} software \citep{lightkurve}. We found rather flat light curves with no obvious photometric variability higher than $\sim$5\,\%~(confidence of 3 $\sigma$). We conclude that with the data in hand, the stellar activity scenario cannot be discarded.

A third scenario is the presence of giant planets around HE\,0107$-$5240 with orbital periods of less than a few hundred days, which are the ones capable of imposing velocity amplitudes of the order of several tens\,m\,s$^{-1}$ on their parent star. We explored this hypothesis by fitting one and two Keplerian models with null eccentricity to the ESPRESSO velocity residuals finding that the two-planet model with velocity amplitudes of $\approx$30 and 40\,m\,s$^{-1}$ is preferred by the Bayesian statistics. The two planets would have minimum masses of $\approx$0.5 and 1.0\,M$_{\rm jup}$ and orbital periods below 200\,d. However, there is neither a sufficient number of radial velocity measurements nor there is the cadence of the spectroscopic data adequate for a in-depth planetary analysis. Studying further this scenario is beyond the scope of this paper. 

We remark, however, that should the presence of giant planets orbiting the hyper metal-poor star HE\,0107$-$5240 be confirmed, this would challenge the core accretion theory of planet formation because heavy elements are needed to cool down the gas of the protoplanetary disk and form the dust grains that settle into the midplane and coagulate to create the planetesimals \citep{johnson12}. 

A fourth scenario is actually the mixture of two or more scenarios above. More ESPRESSO data are planned to rule out any hypotheses.

\section{Improved Chemistry  of HE\,0107$-$5240 }\label{sec:chemistry}

\begin{figure*}
\begin{center}
{\includegraphics[width=180 mm, angle=0]{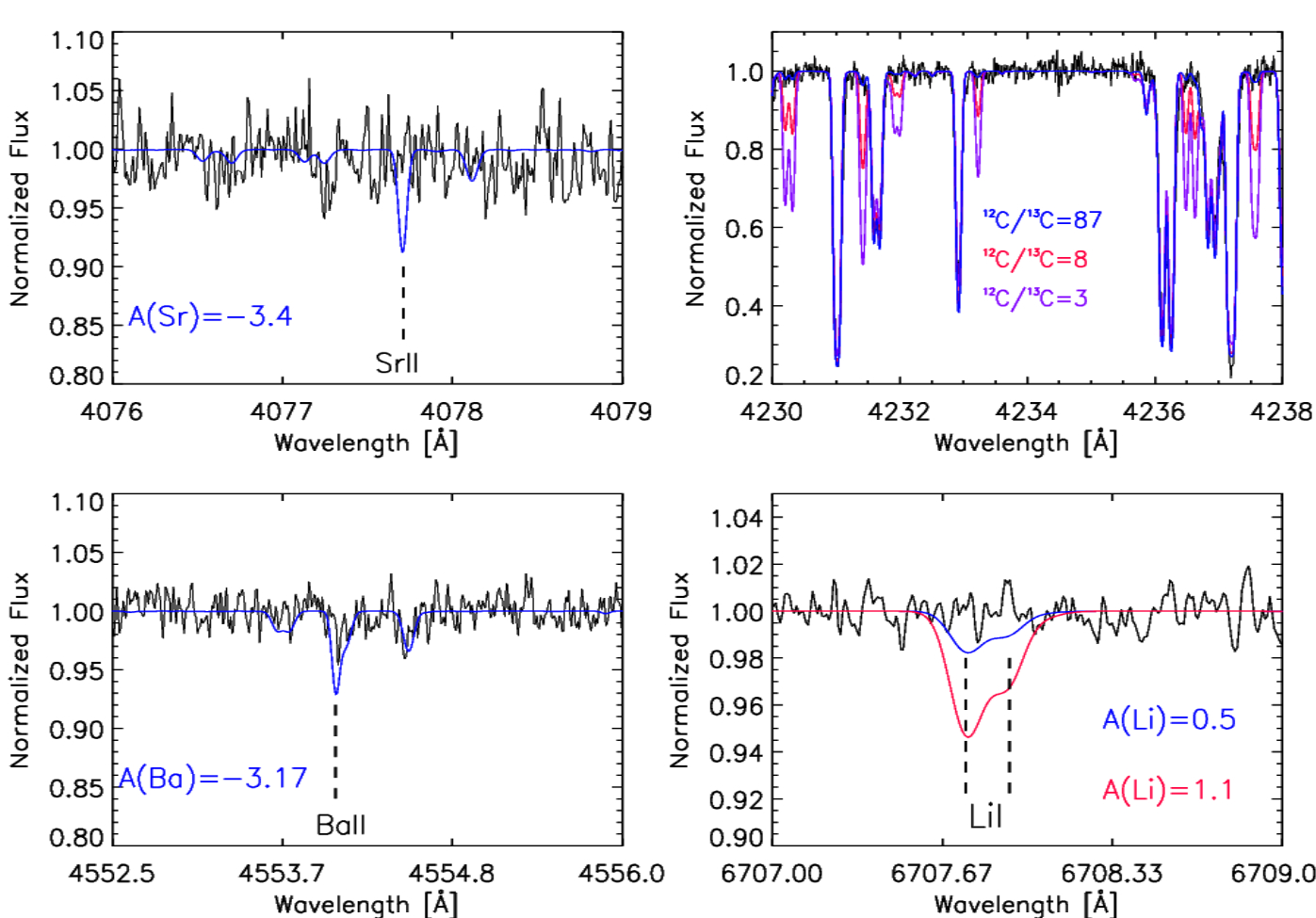}}
\end{center}
\caption{A narrow region of the ESPRESSO combined spectrum of  HE\,0107$-$5240 around the \ion{Sr}{ii} line at 4077.7\,\AA\,\, (upper-left), the G$-$band (upper-right), the \ion{Sr}{ii} line at 4554.0\,\AA\,\, (lower-left), and the Li doublet at 6707.8\,\AA\,\, (lower-right). Derived values and adopted upper limits are shown in blue. Remarkably, at the ESPRESSO resolution the asymmetry in the profile of Li doublet is clear and the upper limit ($\rm A(Li)<0.5$ is much more constraining than the UVES value from \citet[][]{chris04} ($\rm A(Li)<1.1$) in red.
}

\label{fig:chemistry}
\end{figure*}

The high quality of the combined spectrum shown in Fig. \ref{comp} allow us to perform detailed chemical analysis beyond that existing works about HE\,0107$-$5240 in the literature. The stellar parameters of this star are of T$_{\rm eff} =5200$\,K, $\log g =2.2$ and microturbulence of v$_{mic}= 2.2$\,\kms\,\, are taken from \citet{chris04}. We computed different stellar models by using the {\tt SYNTHE} code \citep{kur05, sbo05} along with the {\tt ATLAS9} model atmosphere code. In table \ref{tab3} we measured 21 \ion{Fe}{i} lines from ESPRESSO data (See Table \ref{tab3}) leading  to a $\rm A(\ion{Fe}{i})=1.96\pm0.13$. By assuming $\rm A(\ion{Fe}{i})_{\odot}=7.52$ \citep{caffau11b,lodders19p}, we end up with $\rm [Fe/H]=-5.56\pm0.13$ while the preferred metallicity originally reported by \citet{chris04} was $\rm [Fe/H]_{LTE}=-5.39\pm0.20$. In Table \ref{tab3} the lines and the derived abundances are reported and compared with those of \citet{chris04}. The general abundances are confirmed and some more lines are observed. In general, the derived abundances are slightly lower than those originally derived. Few significant improvements are obtained.  The measured $^{12}$C/$^{13}$C ratio implies that the star is unmixed (i.e.,  chemically unmixed, \citealt[][Fig. 3c]{spite06}). The  previous upper limits were  $\rm [Sr/Fe] \leq -0.52$ and $\rm [Ba/Fe] \leq +0.82$ \citet{chris04} are now improved into $\rm [Sr/Fe] \leq -0.76$ and $\rm[Ba/Fe] \leq +0.20$. In particular, the latter now allows to classify HE\,0107$-$5240 as a CEMP-{\it no} with confidence, which requires $\rm [Ba/Fe]< 1.0$. The upper limit on the lithium is now determined at A(Li)$<$ 0.5. In Fig. \ref{fig:chemistry}, we show our new $^{12}$C/$^{13}$C measurement and the more constraining upper limits on Li, Sr, and Ba.

 \begin{table}
\caption{Abundances of HE\,0107$-$5243 from the  ESPRESSO spectrum. In the 6th column the abundances from \citet{chris04} are given for comparison. In the column for comments some possible blends are indicated.}
\label{tab3}
\scriptsize
\hspace*{0cm}\resizebox{1.\linewidth}{!}{
\begin{tabular}{lcccccccl}
\hline
\hline
\multicolumn{1}{c}{{Specie}} &
\multicolumn{1}{c}{{Ion}} &
\multicolumn{1}{c}{{$ \lambda (\AA)$}} & 
\multicolumn{1}{c}{A(X)$_{\odot}$$^{1}$} & 
\multicolumn{1}{c}{A(X)} & 
\multicolumn{1}{c}{A(X)$_{\rm Ch04}$} &
\multicolumn{1}{c}{{R.I.}} & 
\multicolumn{1}{c}{{[X/Fe]$^{2}$}} & 
\multicolumn{1}{c}{Notes} \\
\hline
Li  & 3.00  &  6707.76  & 1.04 & <0.50  & <1.1  & --    &--&			\\
C & 6.00  &  G-band   & 8.58 & 6.90   &  6.81 &  --    &+3.88&			\\
$^{12}$C/$^{13}$C & 6.00  &  G-band   &  91.4    &  87  & $\sim60$      & --     &--&       \\
N   & 7.00  &  CN       & 8.00 & 5.00   & 5.22  &--      &+2.56&       \\
Na  &11.00  &  5889.951 & 6.28 & 1.62   & 1.83  &--      &+0.90&          \\
Na  &11.00  &  5895.924 &      & 1.62   & 1.90  &--      &+0.90&            \\
Mg  &12.00  &  3829.355 & 7.50 & 2.15   &--       & 0.47 &+0.21&           \\	 
    &12.00  &  3832.304 &      & 2.22   &--       & 0.34 &+0.29&  	\\
    &12.00  &  3838.292 &      & 2.09   &--       & 0.23 &+0.15&  	\\ 
    &12.00  &  5167.321 &      & 2.33   &--       & 0.88 &+0.39&  	\\
    &12.00  &  5172.684 &      & 2.31   & 2.38  & 0.65 &+0.37&  	\\
    &12.00  &  5183.604 &      & 2.31   & 2.45  & 0.53 &+0.37&  	\\ 
Al  &13.00  &  3961.520 & 6.41 & <0.90  & <0.93 & 0.58 &<+0.05&				\\
K   &19.00  &  7698.964 & 5.10 & <1.22  &--       & --     &<+1.68&				\\  
Ca  &20.01  &  3933.663 & 6.34 & 1.39   & 1.44  & 0.15 &+0.61&				\\
    &20.01  &  3968.469 &      & 1.47   &--       & 0.15 &+0.69&				\\
    &20.00  &  4226.728 &      & 0.78   & 0.99  & 0.63 &+0.00&          G-band 		\\
Cr  &24.00  &  5204.511 & 5.62 & <1.07  &--       & 0.99 &<+1.01&				\\  
    &24.00  &  5206.023 &      & <0.87  &--       & --     &<+0.81&				\\  
    &24.00  &  5208.425 &      & <0.67  &--       & 0.97 &<+0.61&				\\  
Fe  &26.00  &  3820.425 & 7.52 & 1.71   & 1.91  & 0.23 &-0.25&				\\
    &26.00  &  3824.444 &      & 2.01   & 2.30  & 0.49 &+0.05&				\\
    &26.00  &  3825.881 &      & 1.73   & 1.99  & 0.32 &-0.23&				\\
    &26.00  &  3827.822 &      & 1.93   &--       & 0.70 &-0.03&				\\
    &26.00  &  3834.222 &      & 1.85   &--       & 0.51 &-0.11&       H wing 		\\
    &26.00  &  3859.911 &      & 1.79   & 2.08  & 0.19 &-0.17&				\\
    &26.00  &  3878.573 &      & 2.04   & --      & 0.57 &+0.08&          CN 		\\
    &26.00  &  3886.282 &      & 1.98   & --      & 0.36 &+0.20&          CH		\\
    &26.00  &  3899.707 &      & 2.13   & --      & 0.66 &+0.19&				\\
    &26.00  &  3922.912 &      & 2.18   & 2.24  & 0.71 &+0.22&				\\
    &26.00  &  3930.297 &      & 2.11   & --      & 0.70 &+0.15&				\\
    &26.00  &  4045.812 &      & 1.90   & 2.03  & 0.52 &-0.06&				\\
    &26.00  &  4063.594 &      & 1.92   & 2.00  & 0.69 &-0.04&				\\
    &26.00  &  4071.738 &      & 1.97   & 2.17  & 0.76 &+0.01&				\\
    &26.00  &  4143.868 &      & 1.95   & --      & 0.89 &-0.01&				\\
    &26.00  &  4202.029 &      & 2.14   & --      & 0.92 &+0.18&  			\\
    &26.00  &  4250.787 &      & 1.88   & --      & 0.94 &-0.08&         G-band  		\\
    &26.00  &  4325.762 &      & 1.91   & --      & 0.76 &-0.05&          G-band  		\\
    &26.00  &  4383.545 &      & 1.93   & --      & 0.58 &-0.04&           G-band  		\\
    &26.00  &  5269.537 &      & 2.08   & 2.20  & 0.91 &+0.12&				\\
    &26.00  &  5328.039 &      & 2.08   & --      & 0.94 &+0.12&				\\
Sr  &38.01  &  4077.709 & 2.92 & <-3.40 &<-2.83 & 0.48 &<-0.76&			\\
    &38.01  &  4215.519 &      & <-3.40 &<-2.53 & 0.65 &<-0.76&			\\
Ba  &56.01  &  4554.029 & 2.19 & <-3.17 & --      & 0.91 &<+0.20&				\\
    &56.01  &  4934.076 &      & <-2.87 &<-2.33 & 0.95 &<+0.50&				\\
\hline
\hline
\end{tabular}}
\tablefoot{\tablefoottext{1}{Solar abundances from \citet{lodders19p}}\\
  \tablefoottext{2}{Chemical ratios, [X/Fe], are based in our derived LTE metallicity: $\rm [Fe/H]=-5.56\pm0.13$}}
\end{table}

\section{Speckle observations of HE\,0107$-$5240}\label{speckle}

 HE\,0107$-$5240 is a binary system but the mass and nature of the companion are uncertain. A giant is excluded since it would outshine any unevolved companion  \citep{boni20}. \citet{arentsen19}  suggested that the companion is a white dwarf which in its AGB phase transferred mass to the currently observed giant. Unfortunately, the spectral energy distribution (SED) of HE\,0107$-$5240 does not provide useful indications for the companion star. HE\,0107$-$5240 is detected in $NUV$ filter ($\lambda_{\rm eff}=213.6$\,nm but not in the $FUV$ filter ($\lambda_{\rm eff}=153.9$\,nm  of GALEX survey  \citep{bianchi2011}. \citet{venn2014ApJ...791...98V}  explored the spectral energy distribution from 444\,nm to 22080\,nm claiming a marginal detection of an IR excess in the WISE  $W3$  band ($\lambda_{\rm eff}=11560$\,nm).
 
In order to set further limits on the nature of the companion star, we exploit speckle imaging on the Gemini 8-m telescope which provides an angular resolution equal to the diffraction limit of the telescope and fairly deep image contrast levels. We  observed HE\,0107$-$5240  on 2019 September 16 UT and 2019 October 8 UT using the Zorro speckle instrument mounted on the 8\,m Gemini-South telescope located on Cerro Pachon, Chile \citep{SH21,HF21}. Zorro uses a dichroic to split the optical light at 674\,nm and obtains simultaneous blue and red images onto two Andor EMCCD detectors. 
Both observations of HE\,0107$-$5240 gave consistent results, but the October observations occurred under better seeing, thus the image contrast achieved was slightly larger.
The 8 Oct 2021 UT observation consisted of a total on-source time of 15 minutes bracketed by observations of point-spread function (PSF) standard stars HR 0242 and HR 0667. The observation collected 1200 images of 60\,ms which are then subjected to Fourier analysis techniques and used to search for close companions, to determine their properties, and provide reconstructed high resolution images \citep[see][]{howell2011AJ....142...19H}. 

\begin{figure}
\begin{center}
{\includegraphics[width=85 mm, angle=0]{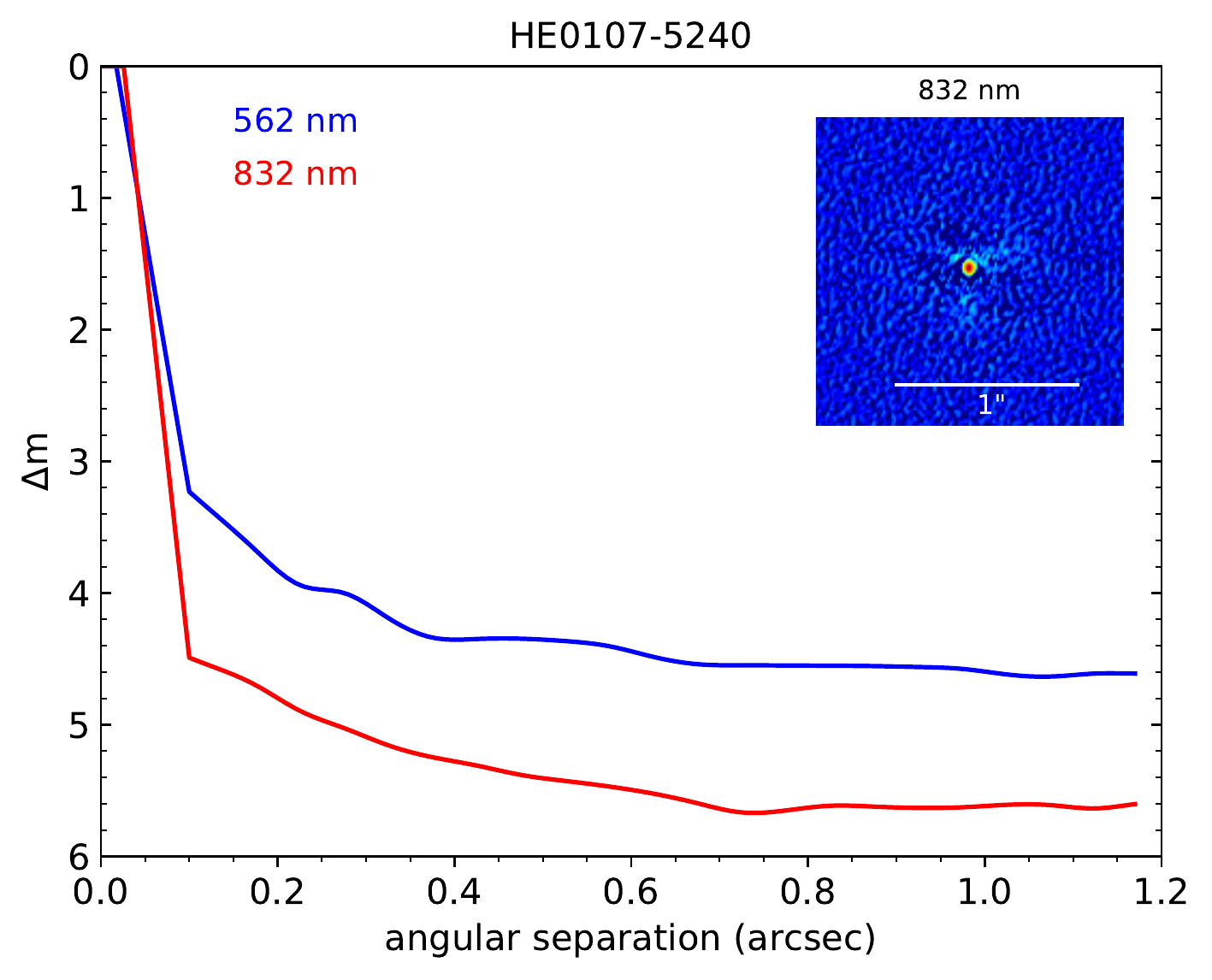}}
\end{center}
\caption{The panel presents our final 5$\sigma$ contrast curves, in 562 nm and 832 nm, that is the limiting magnitude reached in each filter as a function of angular distance from the diffraction limit out to 1.2 arcsec. The insets shows the 832 nm speckle interferometric reconstructed image in which the brighter primary component is at the center. }
\label{fig:specklef}
\end{figure}

Figure \ref{fig:specklef} shows the achieved image contrast curves (differential magnitude as a function of the projected angular separation to the central star) for the Oct 2019 HE\,0107$-$5240 observation. The angular limits extend from the 8-m diffraction limit, near 20\,mas, out to the speckle correlation limit near 1.2\,arcsec. No companion star was detected with differential magnitudes of less than 4 mag (562 nm) and 5.8 mag (832 nm) around HE\,0107$-$5240. Any fainter source would have not been seen by the Speckle observations. This is not surprising giving the distance to HE\,0107$-$5240 of $\approx$ 11\,Kpc, as the angular limits of 20\,mas to 1.2\,arcsec correspond to a linear distance of about 220\,AU to 13,200\,AU from the primary. The inner working angle distance is not far away from the approximate distance of a secondary clump ($\approx$ 150\,AU) in the simulations of protostar formation from clouds with primordial chemistry of \citet{becerra2015MNRAS.446.2380B} but much higher than the few AU of the kinematical motion described in Sec \ref{sec:anal}.

\section{Follow-up of CEMP-{\it no} stars with $\rm [Fe/H]<-4.5$}\label{sec:others}
At the time of writing there are 12 CEMP-{\it no} stars and 2 Carbon-normal stars with $\rm [Fe/H] < -4.5$.  As part of the same GTO ESPRESSO program, besides HE\,0107$-$5240, we targeted 7 more relatively bright CEMP-{\it no} stars living in the most metal-poor regime.
The remaining 6 objects already identified as primitive stars with $\rm [Fe/H]<-4.5$ have not been observed in this program for various reasons. For the sake of completeness we briefly comment on them:
\begin{itemize}
\item SDSS\,J1035+0641 (G$_{mag}=18.4$, $\rm [Fe/H]<-5.3$), and SDSS\,J1742+2531 (G$_{mag}=18.6$, $\rm [Fe/H]=-4.8$) were discovered by \citet{boni15,boni18} and are CEMP stars too faint to be observed with ESPRESSO with a reasonable amount of time.
\item SDSS\,J1029+1729 (G$_{mag}=16.5$, $\rm [Fe/H]=-4.73$, \citealt{caff11}), and Pristine$\_$221.8781+9.7844 (G$_{mag}=16.2$, $\rm [Fe/H]=-4.66$, \citealt{sta18}) show no CH absorptions and only an upper limit for carbon abundance have been provided. This carbon-normal population was out of the scope of this project.
\item SDSS\,J0929+0238 (G$_{mag}=17.9$, $\rm [Fe/H]=-4.97$, \citealt{boni15}) has been proved to be not a binary but a multiple CEMP system \citep{caff16}.
\item SDSS\,J0815+4729 (G$_{mag}=16.7$, $\rm [Fe/H]=-5.5$, \citealt{agu18I,jon20}) is an extremely carbon enhanced star but  not visible from Paranal.
\end{itemize}

We also note that for our 8 stars observed with ESPRESSO the Renormalised Unit Weight Error (RUWE) published by Gaia EDR3 \citet{gaiaidr3} is around 1 in all cases (See Table \ref{tab:par}) and much below 1.4 anyway. Therefore, no binarity could be detect by means of RUWE data.  In the following  we show the results from our follow-up ESPRESSO observations of the other 7 CEMP-{\it no} stars with $\rm [Fe/H] < -4.5$.

\subsection{SDSS\,J0023+0307}
Firstly reported by \citet{agu18II} and subsequently observed at high-resolution \citep{agu19a, fre18}, J0023$-$0307 is a $\rm [Fe/H]<-6.1$ dwarf star with no iron lines detected yet. We visited J0023$-$0307 three times during Oct-Dec 2019 with approximately one month interval between observations done with the same setup. This star is faint ( G$_{mag}=17.6$) and therefore only the prominent \ion{Mg}{i}\,b were marginally detected. Consequently, the CCF can be computed only with a synthetic template and the errors are large ($\sim150\,\rm m\,s^{-1}$). All the three $v_{rad}$ measurements obtained with ESPRESSO are consistent with literature values ($-195.5\pm1.0$ and $-194.6\pm1.2$\,\kms\,\, for \citet{agu19a} and \citet{fre18} respectively. Therefore no evidence for radial velocity variation is shown by the measurements. We report the ESPRESSO values in Table \ref{table:rvs} but not including this object in Fig. \ref{fig:others} for clarity.

\subsection{HE\,0233$-$0343}
This turn-off star with  $\rm [Fe/H]=-4.68$ \citep{han14} and  G$_{mag}=15.3$ was observed four times between Aug-Oct 2019. The G-band around 4330\AA\,\, is detected  in the ESPRESSO spectra and was used to cross-correlate with the natural template. As a result, the four observations with an average value of ~52.107\,\kms\,\, are compatible at 2\,$\sigma$.  \citet{han14} reported $v_{rad}=64$\,\kms\,\, from high-resolution UVES data. To confirm this hint of variability we retrieved the UVES data from the ESO archive -program ID 69.D-0130(A). Following a similar approach to the ESPRESSO data, we derived from three UVES existing spectra $v_{rad}=52.0\pm0.6$, $51.7\pm0.6$, and $51.4\pm0.6$\,\kms\,\, respectively. This completely agrees with the ESPRESSO measurements summarised in Table \ref{table:rvs}. We speculate that the measure of \citet{han14} does not account for the   barycentric correction which is  between V$_{bar}$=-10.5 and -12.0\,\kms\,\,, that would explain the discrepancy.  Further observation would be required to verify longer-term binarity.

\subsection{SMSS\,0313$-$6708}
At the time of writing SMSS\,0313$-$6708 is the most iron-poor star ever known with $\rm [Fe/H]<-7.3$ and G$_{mag}=14.5$. This target was visited 10 times in the period Aug 2019$-$Feb 2021. Three observations  were repeated within the same night
but removed from the $v_{rad}$ series since is significantly lower quality than the other observations.
The first group of data corresponding to 2019 seem consistent with no variation while the measurements in 2012 provide light hints of decreasing behaviour (See Table \ref{fig:others}).  All together, the measurements are consistent with no variation at 2$\sigma$ but further observations could provide a more complete picture if were this star a large period binary system. The chemistry associated with these observations will be published in a separate article.

\subsection{HE\,0557$-$4840}
Identified by \citet{nor07}, this evolved UMP star with $\rm [Fe/H]=-4.75$ and  G$_{mag}=15.2$ was observed with ESPRESSO  in three different nights in the period  Oct 2019-Jan 2020. Due the weakness of CH absorptions in the G-band, the CCF was built around the \ion{Mg}{i}\,b region and the  $v_{rad}$ measurements are fully compatible with an average of 212.2\,\kms\,\, (See Fig. \ref{fig:others}). Other measurements from the bibliography included in \citet{nor07} and \citet{arentsen19} report a similar mean value though with larger  error bars at the level of 1\,\kms. 

\subsection{SDSS\,J1313$-$0019}
SDSS\,J1313$-$0019, G$_{mag}=16.4$, was discovered by \citet{alle15} and lately studied by \citet{fre15} and \citet{agu17II}. According to the high-resolution data from Mike at Magellan \citep{fre15}, this evolved star is $\rm [Fe/H]=-5.00$, highly carbon-enhanced $\rm [C/Fe]=+3.0$, and shows $v_{ rad}=274.6$\,\kms\,\, (no error bars were reported). We observed it just once in March 2022 -6 years later- giving as a result $v_{rad}=273.98\pm0.05$\,\kms. The observed difference ($\sim620\,m\,s^{-1}$) is clearly smaller than most optimistic uncertainty from Mike observations ($1-2$\,\kms). Furthermore, by cross matching this object with the \textit{Survey of Surveys} database \citep{Tsantaki22}, we found $v_{rad}=273.054\pm4.498$\,\kms\,\, from SEGUE data, in perfect agreement with those from high-resolution observations. There are, however, other measurements from low-resolution observations reported by \citet{alle15}, $v_{rad}=268\pm6$\,\kms\,\, (July 2008, SEGUE), and  $v_{rad}=242\pm4$\,\kms\,\, (March 2014, BOSS). While both SEGUE values are consistent with Mike and ESPRESSO data, the BOSS measurement (which is closer in time) seems to be incompatible. That fact allowed \citet{alle15} and \citet{fre15} to suggest that SDSS\,J1313$-$0019 could be a binary system. However, the large uncertainty given by the low-resolution data and the fact that both high-resolution measurements are largely compatible do not allow us to claim for $v_{rad}$ variability. Alternatively, the data in hand does not preclude completely some binarity behaviour, more ESPRESSO measurements are required.

\subsection{HE\,1327$-$2326}
Firstly studied by \citet{fre05}, this HMP star with $\rm [Fe/H]=-5.60$ and  G$_{mag}=13.4$ was observed 9 times between May 2019 and March 2022. Despite the observations span over three years no evidence of variability is found. The second observation (See Fig. \ref{fig:others}) deviates from  the main value by more than 2$\sigma$ but the quality of this spectrum was much lower than the others due to poor observing conditions. As in the case of SMSS\,0313$-$6708, the chemical analysis of the combined spectrum will be published separately in a forthcoming paper.

\subsection{SMSS\,1605$-$1443}

\begin{figure}
\begin{center}
{\includegraphics[width=90 mm, angle=0]{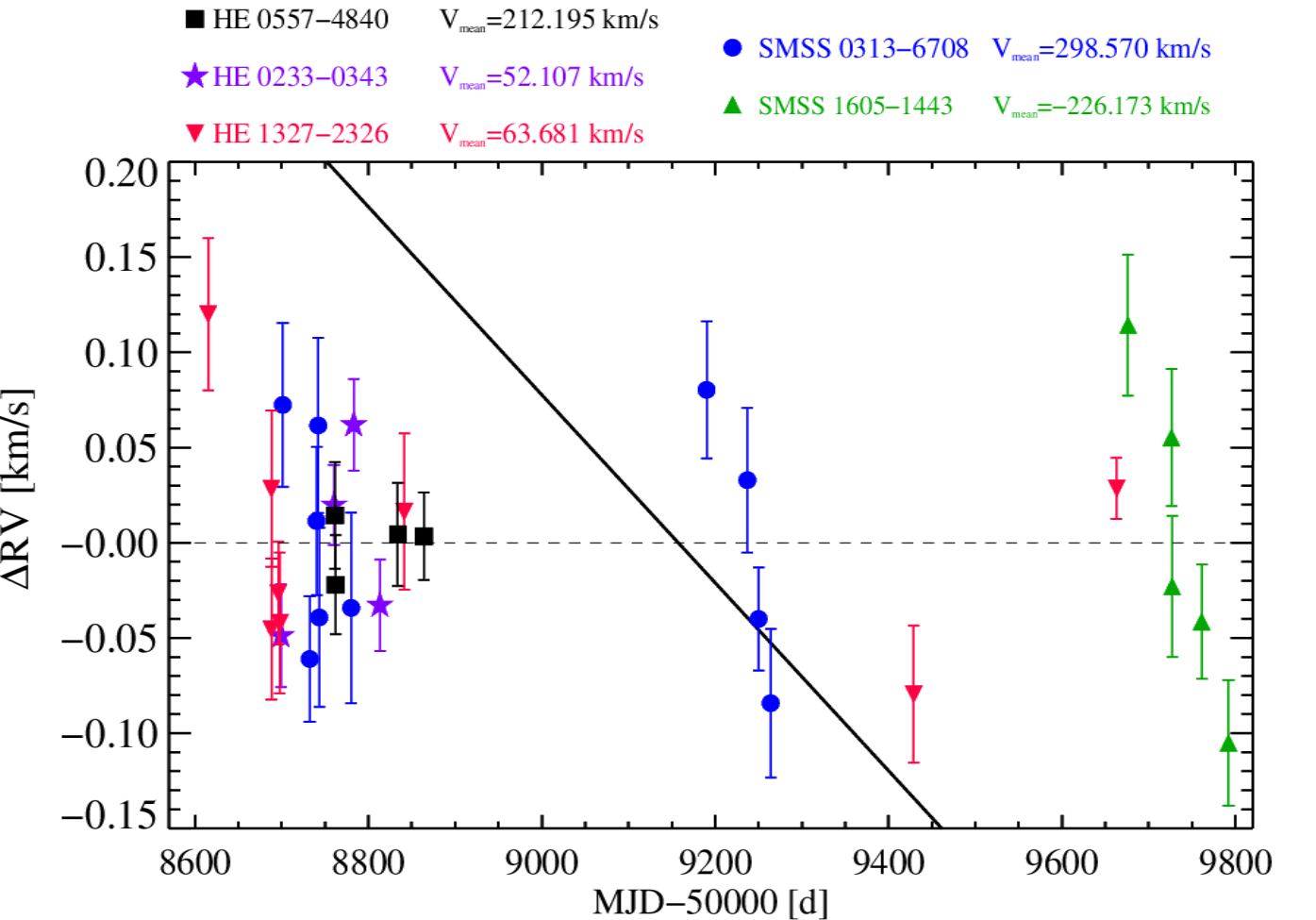}}
\end{center}
\caption{A summary of  the  $v_{rad}$ ESPRESSO measurements discussed in Section \ref{sec:others} in which the baseline is the average for each star for display purposes. The stars are shown with different colours and symbols: black square is  HE\,0557$-$4840 with a mean velocity taken at 212.195 \kms; violet stars is HE\,0233$-$0343 with ($v_{rad})_{mean}=$ 52.107 \kms; red downwards triangles is HE\,1327$-$2326 with ($v_{rad})_{mean}=$ 63.681  \kms; blue dots is SMSS\,0313$-$6708 with ($v_{rad})_{mean}=$ 298.560 \kms; green upwards triangles SMSS\,1605$-$1443 ($v_{rad})_{mean}=$ -226.173 \kms.  Errors are of 1 $\sigma$. For reference, the black line crossing the panel is the average variation of HE\,0107$-$5240 in the same interval.}
\label{fig:others}
\end{figure}

SMSS\,16054$-$1443 is a recent MMP halo giant discovered by \citet{nordlander2019MNRAS.488L.109N} with the SkyMapper telescope.  With  G$_{mag}=15.4$ and a metallicity of $\rm [Fe/H]= -6.2$ (1D-LTE), it is the lowest abundance of iron ever detected in a star. The star is strongly carbon-enhanced, [C/Fe] = 3.9,  and has no significant s$-$ or r$-$process enrichment, [Sr/Fe] $<$ 0.2 and [Ba/Fe] $<$ 1.0 being very likely a CEMP-{\it no} star. The authors in \citet{nordlander2019MNRAS.488L.109N} reported $v_{rad}=-224$\,\kms\,\, from observations with Mike at Mallegan telescope in September 2018, although no error bars were provided. Based in our experienc,e a reasonable uncertainty for that measurement could be of the order of $1-2$\,\kms. 43 months later we detected a $\sim2$\,\kms difference from 5 ESPRESSO measurements  ($v_{rad})_{mean}=$ -226.173\,\kms\,\, (see Table \ref{table:rvs}). Therefore, we can expect a decrease rate of $\sim1.6\,\rm m\,s^{-1}\,$d$^{-1}$. Our observations span $\sim115$\,d and show a decrease of $\sim219\,\rm m\,s^{-1}$ (see Fig. \ref{fig:others}), giving a rate of approximately $\sim1.9\,\rm m\,s^{-1}\,$d$^{-1}$ which is about consistent with the rate derived comparing Mike and ESPRESSO. We find this result convincing and tentatively propose that SMSS\,16054$-$1443 could be another system similar to HE\,0107$-$5240 but varying 3 times faster. More ESPRESSO observations are required to confirm this hypothesis and calculate orbital parameters.

\section{Discussion}\label{sec:discu}

\subsection{Binarity of our CEMP-{\it no} sample}
We have probed 8 of the most extreme metal poor CEMP-{\it no} stars for binarity by means of very accurate radial velocity measurements obtained with ESPRESSO and we only found clear evidence for binarity only in the case of HE\,0107$-$5240.
For this star \citet{suda04} invoked  mass transfer from a companion AGB in a similarly to what is believed to happen in CEMP-{\it s} stars (i.e., stars with $\rm [Ba/Fe]>1$).  Moreover,  \citet{arentsen19} suggested the presence of a radial velocity variation by comparison of the UVES observations with some observations taken at the  High-Resolution Spectrograph (HRS) on the SALT Telescope. They noted that  the radial velocity of HE\,0107$-$5240 was almost 4\,\kms\ larger than at the discovery time. On the basis of possible binarity \citet{arentsen19} suggested that the enhancement in Carbon could have been the result of mass transfer from the companion star.
The period derived here for HE\,0107$-$5240 is so long and the minimum mass of the companion so low  that very unlike there is any interaction or mass exchange between the two stars of the system. 
The result for HE\,0107$-$5240 and for the other 7 stars is therefore quite consistent with the CEMP-{\it no} being either single or in  binary systems with low mass companions as found by \citet{han16I,han16III} in a sample of more metal rich CEMP-{\it no} stars.

\subsection{The stellar structure of HE\,0107$-$5240}
 HE\,0107$-$5240 lies in the position of the LRGB in the \teff-\logg\,\, diagram  \citep{mucciarelli2022arXiv220310347M}, and  has not experienced the mixing episode that occurs at the Red Giant Branch bump, and therefore the chemical composition is expected to be that of the natal cloud.
Thus, the metallicity is so low that  several  or even just one SNe are enough to yield all the metals  measured in the stellar atmospheres  \citep{ume03,boni03, lim03}. The observed carbon abundances in CEMP-{\it no} stars is believed to come from faint supernovae of energy of $10^{51}$\,erg,  where fallback and mixing  mechanisms played a role \citep{ume03}.  \citet {boni15} proposed an alternative explanation invoking double source: lighter elements, typically CNO, synthesised by faint SNe but heavier elements, some $\rm \alpha's$ included, by classical core-collapse SNe.
For HE\,0107$-$5240 we provide more stringent upper limits for the neutron capture elements and a first measure of the $^{12}$C/$^{13}$C. \citet{hansen2015ApJ...807..173H} found evidence of a {\it floor} in the absolute Ba abundance of CEMP-{\it no} stars at the abundance level of A(Ba) $\approx$ -2.  The existence of such a plateau was considered a specific feature of the CEMP-{\it no} stars and related to their progenitors. Both \textit{spinstar} models \citep[i.e., metal-poor stars rotating fast][]{mey06,choplin2018A&A...618A.133C} and mixing and fall back models \citep{ume03,Nomoto2013} predict the production of an important amount of neutron capture elements which apparently are not seen in HE\,0107$-$5240.  This is generally interpreted as a minimum amount of Ba produced by the \textit{spinstar}s \citep{cescutti2016A&A...595A..91C}. 
 The present upper limit of $\rm A(Ba)<-3.17$ derived here is  more than an order of magnitude below the level of the suggested floor value and provides evidence  of a much reduced n-capture element production by the early generation of stars. 
 
 \subsection{The isotopic ratio $^{12}$C/$^{13}$C}
In HE\,0107$-$5240 we have been able to measure the  $^{12}$C/$^{13}$C ratio. $^{12}$C is a primary product of stellar nucleosynthesis and is formed in the triple-$\alpha$ process during hydrostatic helium burning. The stable $^{13}$C isotope is  produced in the hydrogen-burning shell when the CN$-$cycle converts $^{12}$C into $^{13}$C. These processes, which occur mainly in intermediate massive stars which develop as AGB \citep{wannier80}, lead to a CN$-$cycle equilibrium ratio of about  $^{12}$C/$^{13}$C  $=\sim$3.4. Large quantities of $^{13}$C could also be produced by low-metallicity massive and fast-rotating stars which should mostly affect the early chemical evolution  \citep{Meynet2006A,Limongi2018,maeder2015A&A...576A..56M}. \citet{Meynet2006A}  showed that the  isotope composition in the stellar wind is characterised by very low values of $^{12}$C/$^{13}$C   $\sim 3-5$. 
Due to the contribution of relatively long-lived stars the  chemical evolution models of the Galaxy predict a decrease of the isotopic ratio $^{12}$C/$^{13}$C with time, and an increase with galactocentric distances \citep{Romano2003MNRAS.342..185R}.  In fact,  the  photospheric solar ratio  $^{12}$C/$^{13}$C$ = 95\pm5$ \citep{Asplund2006NuPhA.777....1A}  is higher than the   value found in local molecular clouds   ($^{12}$C/$^{13}$C$\sim 60-70$, \citealt{goto03, milam05}), and is significantly higher than the at measured in the  Galactic  Center by  \citet{Wilson1992}, $^{12}$C/$^{13}$C$=25\pm 5$.

At high redshift,  carbon abundance had  been measured in the metal-poor damped Ly$\alpha$ systems (DLA). \citet{Welsh2020} reported a bound on the carbon isotope ratio  $\rm ^{12}C/^{13}C > 2.3 \, (2\sigma)$ in a DLA system at $z=2.34$.  They found also tentative evidence that the \emph{most} metal-poor DLA population exhibits somewhat higher [C/O] values at redshift $z\lesssim3$ with a potential upward trend in the ratio going towards lower redshift. They argued that  the elevated [C/O] ratios at $z\lesssim3$ might be a signature of enrichment from the first metal-enriched low- and intermediate-mass stars. 
 At somewhat lower redshift there are few  determinations of the carbon isotopic ratio.  \citet{Muller2006} obtained a $^{12}$C/$^{13}$C ratio of 27 $\pm$ 2 in the Galaxy at z$_{abs} =0.89$ by using the \ion{C}{i} transitions.  \citet{Levshakov2006} obtained $^{12}$C/$^{13}$C$> 80$ in the z$_{abs}$ = 1.15 DLA towards HE\,0515$-$4414  implying a very low  $^{13}$C abundance.  \citet{Carswell2011}  in the z$_{abs}$ = 1.7764  DLA system toward  J1331\,+\,170, which is also the  system  we are studying here by means of new ESPRESSO observations, derived $^{12}$C/$^{13}$C  $>$ 5 (2$\sigma$). 
 
 \cite{hansen2015ApJ...807..173H,norris2013ApJ...762...28N} measured the $^{12}$C/$^{13}$C  in a sample of CEMP stars. Most of them have low $^{12}$C/$^{13}$C as a result of high $^{13}$C, but they also show [C/N] $<$ 0 which is   evidence of internal mixing according to \citet{spite06}. HE\,0107$-$5240 shows [C/N] $>$ 0 and therefore is unmixed. Consequently, the new measure of $^{12}$C/$^{13}$C  $=87\pm6$ implies that the low $^{13}$C is pristine. This suggests that the parent generation of stars that polluted the material from which this star is born did not produce any  significant  amount of $^{13}$C. \textit{Spinstar} models \citep{mey06}, which predict production of an important amount of $^{13}$C, are therefore not favoured.

\subsection{The lithium during the HE\,0107$-$5240 evolution}
The upper limit of Li in HE\,0107$-$5240 is also improved with A(Li) $<$ 0.5 with a confidence level higher than $3\,\sigma$. HE\,0107$-$5240 is in the lower red giant branch, LRGB, namely stars  with 2 $\le  \log g \le $ 3 located after the first dredge-up (FDU) and before the RGB-bump luminosity.  Assuming these stars start with the same initial Li abundance,  the present Li abundance depends on how much Li has been  destroyed in the extension of the convection zone during the FDU. \citet{mucciarelli2022arXiv220310347M} found that the majority of LRGB  show a very thin plateau in the Li abundance with $\rm A(Li)= 1.09 \pm 0.07$ and that this results from the stellar dilution in the giant evolution. They also found that a small fraction of stars of about 20\% shows no detectable Li, namely $\rm A(Li) \le 0.7$. For  HE\,0107$-$5240 \citet{mucciarelli2022arXiv220310347M} derived  A(Li) $\le$ 0.7 which we improve slightly in  A(Li) $\le$ 0.5. This bound  places HE\,0107$-$5240  in the  minority of stars without any Li signature. Another LRGB  we are investigating for radial velocity changes, HE\,0557$-$4840, shows also absence of Li with A(Li) $<$ 0.60 \citep{mucciarelli2022arXiv220310347M}.  It is not clear for these stars if they were originally Li poor objects or if they suffered from extra destruction of Li either during the main sequence or the post-main sequence evolution. However, this seems uncorrelated with metallicity since Li has been measured at A(Li) = 0.87 $\pm$ 0.05 also in SMSS\,J0313$-$6708 which, with $\rm [Fe/H]< -7.0$,  is the lowest metallicity star known and, therefore, a similar abundance is expected also in the other more metal rich subgiants. 
Since theoretically there is no known mechanism that can destroy Li in the MS or in the RGB before the RGB-bump, there are  three main scenarios left to explain the metal-poor CEMP-{\it no} stars with no or low Li detection:
\begin{enumerate}
    \item The gas from which the star was formed was Li-free. This may be the case if the star
    was formed from almost pure SN ejecta, with very little dilution by primordial gas.
    \item  Li was destroyed in the pre-main sequence phase \citep{fu2015MNRAS.452.3256F}. This phase is not easily modelled and the question remains in what 
    the pre-MS of these stars was different, with respect to that of other stars that retained Li. In \citet{fu2015MNRAS.452.3256F} the model of the resulting Li is a balance between a pre-main sequence destruction and partial accretion of gas with Li at the primordial value. In general, these stars less massive for a given T$_{\rm eff}$ and such a balance could have been broken.
    \item The star is an evolved  Blue Straggler. \citet{Ryan01} suggested that unevolved metal-poor stars with no detectable Li were ``Blue Stragglers-to-be'', and in fact the Gaia parallaxes have shown that three out of four stars studied by 
    \citet{Ryan01} are indeed ``canonical'' Blue Stragglers \citep{BGaia}. If this were true in the case of HE\,0107-5240, we should
    postulate that the system was initially a hierarchical triple system with an inner pair that eventually coalesced to form a Blue Straggler and an outer companion that is the companion presently revealed by the orbital motion.
\end{enumerate}

\citet{boni18} and \citet{agu19a} studied several dwarf stars with extremely low metallicity, and only 3 out of 8 dwarfs with [Fe/H]$<$ -4.2 have detectable Li close to, but below, the Spite plateau. It is therefore very likely that the high fraction of Li-depleted stars is related to the significant Li destruction by the progenitor  responsible for the high carbon abundances observed in the CEMP-{\it no} stars.
 
\subsection{Consequences on the nature of CEMP-{\it no} stars}
The lack of binaries among the CEMP-{\it no} stars holds important bearings on the star formation of stars from primordial or nearly primordial chemistry.  At the time of the discovery of  HE\,0107$-$5240, the possibility to form a low mass star from almost metal-free gas was not conceived \citep{palla03}. In the conventional model of primordial star formation H$_2$ cooling of metal-free  gas leads to the formation of massive stars with a characteristic stellar mass of $\approx$ 100\,M$_\sun$.

  The discovery of the extremely metal poor CEMP-{\it no} stars led to a revision of the processes for the  formation of low-mass stars.  \citet{bro03} suggested that   atomic lines of neutral oxygen and singly ionised carbon could effectively cool a very metal-poor contracting cloud, provided the abundance of these elements is at the level observed in the CEMP stars. However, the successive discovery  of two extremely metal poor stars which  are not carbon enhanced \citep{caff11, sta18} suggests that  metal-line cooling cannot be the only mechanism capable of leading the formation of a sub-solar mass at very low metallicity.
  Smaller masses of the order of 1\,M$\sun$ may only be obtained by dust cooling. However, recent refinements in the modelling  of the collapse process of primordial star formation have shown that the original clouds are prone to fragmentation \citep{greif2015ComAC...2....3G}. This leads more likely to the formation of a massive central star or a binary system surrounded by some small mass stars. 
In the simulation of \citet{becerra2015MNRAS.446.2380B} of a cooling primordial cloud, they found that the fragmentation of the disc around the initial protostar could generate a second protostellar system with 5$-$10 members with mass of the order of the 1\,M$_\sun$. They also showed that at the end of the simulation the collapse could evolve into a wide binary system with a distance of hundreds of AU. The confirmation of the binarity of HE\,0107$-$5240, which is one of the most metal-poor stars known, shows that the process of star formation in very metal-poor gas could lead to the formation of binaries. However, the majority of single stars show that probably these initial clusters dispersed or did not end up preferentially into binary systems.

\section{Conclusions}\label{sec:conclu}

We summarise here the main conclusions:
\begin{itemize}
    \item Along the time span of 3 years we observed 8 of the most metal poor CEMP-{\it no} stars with ESPRESSO to probe binarity by means of very accurate radial velocities. 
    \item Out of these, only  HE\,0107$-$5240 showed clear evidence for an orbital motion. HE\,0107$-$5240 shows a monotonic decreasing trend in radial velocity at a rate of approximately 0.5\,\rm m\,s$^{-1}\,$d$^{-1}$. The maximum $v_{rad}$ was reached  between Mar 13th, 2012 and Dec 8th, 2014 and the period is constrained at $P_{\rm orb} = 13009_{-1370}^{+1496}$~d. The  $v_{rad}$ for other 6 EMP CEMP-{\it no} stars are consistent with single stars while solid indication of binarity behaviour in SMSS\,1605$-$1443 with an approximate $v_{rad}$ variation of $\sim1.6\,\rm m\,s^{-1}\,d^{-1}$ are found. 
    
    \item  Speckle observations of HE\,0107$-$5240 do not reveal companions at distances greater than 1000\,AU. 
     \item Revised Chemistry  of HE\,0107$-$5240 provided  more stringent upper limits to $\rm [Sr/Fe]< -0.76$  and $\rm [Ba/Fe] < +0.20$   confirming the star is a CEMP-{\it no}. The isotopic ratio $^{12}$C/$^{13}$C is measured for the first time in an unmixed giant and found close to the solar value. The low   $^{13}$C is in contrast to what is expected from a \textit{spinstar} progenitor. Lithium is also improved to a value of  $\rm A(Li)<0.5$ well below the Li plateau at 1.1 found in the Lower Red Giand Branch Stars by \citet{mucciarelli2022arXiv220310347M}, suggesting the star was originally depleted by lithium.  
      \item The fact that there are some binaries among the CEMP-{\it no} group (HE\,0107$-$5240, SDSS\,J0929+0238, and likely  SMSS\,1605$-$1443) indicates that the low metallicity does not inhibit the formation of binaries.
\end{itemize}

\begin{acknowledgements}
The authors wish to thank the Gemini Director for the allocation of engineering time to perform this observation.
DA would like to thank Dr. Megan Bedell for her enjoyable discussion of the ESPRESSO data at the Flatiron Institute.
DA also acknowledges support from the ERC Starting Grant NEFERTITI H2020/808240.
JIGH, CAP, ASM and RR acknowledge financial support from the Spanish Ministry of Science and Innovation (MICINN) project PID2020-117493GB-I00.
JIGH also acknowledges financial support from the Spanish MICINN under 2013 Ram\'on y Cajal program RYC-2013-14875.
A. S. M. acknowledges financial support from the Spanish Ministry of Science and Innovation (MICINN) under 2018 Juan de la Cierva program IJC2018-035229-I.
Zorro was funded by the NASA Exoplanet Exploration Program and built at the NASA Ames Research Center by Steve B. Howell, Nic Scott, Elliott P. Horch, and Emmett Quigley. Gemini Observatory is a program of NSF's NOIRLab, which is managed by the Association of Universities for Research in Astronomy (AURA) under a cooperative agreement with the National Science Foundation. on behalf of the Gemini partnership: the National Science Foundation (United States), National Research Council (Canada), Agencia Nacional de Investigación y Desarrollo (Chile), Ministerio de Ciencia, Tecnología e Innovación (Argentina), Ministério da Ciência, Tecnologia, Inovações e Comunicações (Brazil), and Korea Astronomy and Space Science Institute (Republic of Korea). This work was supported by FCT - Fundação para a Ciência e a Tecnologia through national funds and by FEDER through COMPETE2020 - Programa Operacional Competitividade e Internacionalização by these grants: UID/FIS/04434/2019; UIDB/04434/2020; UIDP/04434/2020; PTDC/FIS-AST/32113/2017 \& POCI-01-0145-FEDER-032113. CJM also acknowledges FCT and POCH/FSE (EC) support through Investigador FCT Contract 2021.01214.CEECIND/CP1658/CT0001.
This work has made use of data from the European Space Agency (ESA) mission {\it Gaia} (\url{https://www.cosmos.esa.int/gaia}), processed by the {\it Gaia} Data Processing and Analysis Consortium (DPAC, \url{https://www.cosmos.esa.int/web/gaia/dpac/consortium}). Funding for the DPAC has been provided by national institutions, in particular the institutions participating in the {\it Gaia} Multilateral Agreement.
This work was financed by FEDER–Fundo Europeu de Desenvolvimento Re-
gional funds through the COMPETE 2020–Operational Programme for Com-
petitiveness and Internationalisation (POCI), and by Portuguese funds through
FCT - Fundação para a Ciência e a Tecnologia under projects POCI-01-0145-
FEDER-028987, PTDC/FIS-AST/28987/2017, PTDC/FIS-AST/0054/2021 and
EXPL/FIS-AST/1368/2021, as well as UIDB/04434/2020 \& UIDP/04434/2020,
CERN/FIS-PAR/0037/2019, PTDC/FIS-OUT/29048/2017.
FPE and CLO would like to acknowledge the Swiss National Science Foundation (SNSF) for supporting research with ESPRESSO through the SNSF grants nr. 140649, 152721, 166227 and 184618. The ESPRESSO Instrument Project was partially funded through SNSF’s FLARE Programme for large infrastructures. MTM acknowledges the support of the Australian Research Council through Future Fellowship grant FT180100194

\end{acknowledgements}
%
%

\bibliography{biblio}

\end{document}